\documentclass[twocolumn]{aastex6}
\pdfoutput=1 
\usepackage{amsmath,amstext}
\usepackage[T1]{fontenc}
\usepackage[figure,figure*]{hypcap}
\usepackage{affils}
\usepackage{hyperref}
\usepackage{microtype}
\usepackage{caption}
\usepackage{subcaption}
\usepackage{xspace}
\usepackage{rotating}
\usepackage[utf8]{inputenc}
\usepackage{CJK}

\def\kms{km\,s$^{-1}$}

\shorttitle{SN~2017ens: Luminous SN~Ic-BL to SN~IIn}
\shortauthors{Chen et al.}

\begin{document}

\title{SN~2017\MakeLowercase{ens}: The Metamorphosis of a Luminous Broadlined Type Ic Supernova into an SN~IIn}

\DeclareAffil{mpe}{Max-Planck-Institut f{\"u}r Extraterrestrische Physik, Giessenbachstra\ss e 1, 85748, Garching, Germany; \href{mailto:jchen@mpe.mpg.de}{jchen@mpe.mpg.de}}
\DeclareAffil{sh}{Department of Physics and Astronomy, University of Southampton, Southampton, SO17 1BJ, UK}
\DeclareAffil{qub}{Astrophysics Research Centre, School of Mathematics and Physics, Queen's University Belfast, Belfast BT7 1NN, UK}
\DeclareAffil{NAOJ}{Division of Theoretical Astronomy, National Astronomical Observatory of Japan, National Institutes for Natural Sciences, 2-21-1 Osawa, Mitaka, Tokyo 181-8588, Japan}
\DeclareAffil{UCD}{School of Physics, O'Brien Centre for Science North, University College Dublin, Belfield, Dublin 4, Ireland}
\DeclareAffil{SSDC}{Space Science Data Center - Agenzia Spaziale Italiana, via del Politecnico, s.n.c., I-00133, Roma, Italy}
\DeclareAffil{FINCA}{Finnish Centre for Astronomy with ESO (FINCA), FI-20014 University of Turku, Finland}
\DeclareAffil{TO}{Tuorla Observatory, Department of Physics and Astronomy, FI-20014 University of Turku, Finland}
\DeclareAffil{stock}{The Oskar Klein Centre, Department of Astronomy, AlbaNova, Stockholm University, SE-106 91 Stockholm, Sweden}
\DeclareAffil{ucsc}{Department of Astronomy and Astrophysics, University of California, Santa Cruz, CA 95064, USA}
\DeclareAffil{ucberkeley}{Department of Astronomy, University of California, Berkeley, CA 94720-3411, USA}
\DeclareAffil{miller}{Miller Senior Fellow, Miller Institute for Basic Research in Science, University of California, Berkeley, CA 94720, USA}
\DeclareAffil{IANCU}{Graduate Institute of Astronomy, National Central University, No. 300, Zhongda Rd., Zhongli Dist., Taoyuan City 32001, Taiwan}
\DeclareAffil{STScI}{Space Telescope Science Institute, 3700 San Martin Drive, Baltimore, MD 21218, USA}
\DeclareAffil{JHU}{Department of Physics and Astronomy, Johns Hopkins University, Baltimore, MD 21218, USA}
\DeclareAffil{INAFcatania}{INAF - Osservatorio Astrofisico di Catania Via Santa Sofia, 78, 95123, Catania, Italy}
\DeclareAffil{INAFnapoli}{INAF - Osservatorio Astronomico di Napoli, Salita Moiariello, 16, I-80131, Napoli, Italy}
\DeclareAffil{IAA-CSIC}{Instituto de Astrofisica de Andalucia (IAA-CSIC), Glorieta de la Astronomia s/n, E-18008 Granada, Spain}
\DeclareAffil{ICRA}{International Center for Relativistic Astrophysics, Piazza della Repubblica 10, I-65122 Pescara, Italy}
\DeclareAffil{AUMRO}{Aalto University Mets{\"a}hovi Radio Observatory, Mets{\"a}hovintie 114, FIN-02540 Kylm{\"a}l{\"a}, Finland}
\DeclareAffil{WaU}{Warsaw University Astronomical Observatory, Al. Ujazdowskie 4, 00-478 Warszawa, Poland}
\DeclareAffil{PITT}{PITT PACC, Department of Physics and Astronomy, University of Pittsburgh, Pittsburgh, PA 15260, USA}
\DeclareAffil{weizmann}{Department of Particle Physics and Astrophysics, Weizmann Institute of Science, Rehovot 76100, Israel}
\DeclareAffil{INAFpadova}{INAF - Osservatorio Astronomico di Padova, Vicolo dell'Osservatorio 5, 35122 Padova, Italy}
\DeclareAffil{UNSW}{School of Physical, Environmental and Mathematical Sciences, University of New South Wales, Australian Defence Force Academy, Canberra, ACT 2600, Australia}
\DeclareAffil{ANU}{Research School of Astronomy and Astrophysics, Australian National University, Canberra, ACT 0200, Australia}
\DeclareAffil{IfA}{Institute for Astronomy, University of Hawaii, 2680 Woodlawn Drive, Honolulu, HI 96822, USA}
\DeclareAffil{MPA}{Max-Planck Institut für Astrophysik, Karl-Schwarzschild Str. 1, D-85748 Garching, Germany}
\DeclareAffil{ESOgarching}{European Southern Observatory, Karl-Schwarzschild-Str. 2, 85748 Garching, Germany}
\DeclareAffil{ESOsantiago}{European Southern Observatory, Alonso de C\'ordova 3107, Casilla 19, Santiago, Chile}
\DeclareAffil{Liverpool}{Astrophysics Research Institute, Liverpool John Moores University, IC2, Liverpool Science Park, 146 Brownlow Hill, Liverpool L3 5RF, UK}
\DeclareAffil{CTIO}{CTIO/NOAO, Casilla 603, La Serena, Chile}
\DeclareAffil{LSST}{LSST, 950 N. Cherry Ave, Tucson, AZ 85719}
\DeclareAffil{DTU}{DTU Space, National Space Institute, Technical University of Denmark, Elektrovej 327, 2800 Kgs. Lyngby, Denmark
}

\affilauthorlist{T.-W.~Chen\affils{mpe},
C.~Inserra\affils{sh}, M.~Fraser\affils{UCD}, T.~J.~Moriya\affils{NAOJ}, P.~Schady\affils{mpe}, T.~Schweyer\affils{mpe}, 
A. V.~Filippenko\affils{ucberkeley,miller}, \\
D. A.~Perley\affils{Liverpool}, A. J.~Ruiter\affils{UNSW,ANU},
I.~Seitenzahl\affils{UNSW,ANU}, J.~Sollerman\affils{stock}, F.~Taddia\affils{stock},  
J. P.~Anderson\affils{ESOsantiago}, R. J.~Foley\affils{ucsc}, A.~Jerkstrand\affils{MPA}, C.-C.~Ngeow\affils{IANCU}, Y.-C.~Pan\affils{ucsc}, A.~Pastorello\affils{INAFpadova}, S.~Points\affils{CTIO},
S. J.~Smartt\affils{qub}, K. W.~Smith\affils{qub}, S.~Taubenberger\affils{ESOgarching,MPA}, 
P.~Wiseman\affils{sh}, D. R.~Young\affils{qub}, 
S.~Benetti\affils{INAFpadova}, M.~Berton\affils{FINCA,AUMRO}, F.~Bufano\affils{INAFcatania}, P.~Clark\affils{qub}, M.~Della Valle\affils{INAFnapoli,IAA-CSIC,ICRA}, L. Galbany\affils{PITT}, A.~Gal-Yam\affils{weizmann}, M.~Gromadzki\affils{WaU}, C. P.~Guti\'errez\affils{sh}, A.~Heinze\affils{IfA}, E.~Kankare\affils{qub}, C. D.~Kilpatrick\affils{ucsc}, H.~Kuncarayakti\affils{FINCA,TO}, G.~Leloudas\affils{DTU},
Z.-Y.~Lin\affils{IANCU}, K.~Maguire\affils{qub}, P.~Mazzali\affils{Liverpool}, O. McBrien\affils{qub}, S. J.~Prentice\affils{qub}, 
A.~Rau\affils{mpe}, A.~Rest\affils{STScI,JHU}, M. R.~Siebert\affils{ucsc}, 
B.~Stalder\affils{LSST}, J. L.~Tonry\affils{IfA}, and P.-C.~Yu\affils{IANCU}
}

\begin{abstract}
We present observations of supernova (SN)~2017ens, discovered by the ATLAS survey and identified as a hot blue object through the GREAT program. The redshift $z=0.1086$ implies a peak brightness of $M_{g}=-21.1$ mag, placing the object within the regime of superluminous supernovae. We observe a dramatic spectral evolution, from initially being blue and featureless, to later developing features similar to those of the broadlined Type Ic SN~1998bw, and finally showing $\sim2000$\,km\,s$^{-1}$ wide H$\alpha$ and H$\beta$ emission. Relatively narrow Balmer emission (reminiscent of a SN~IIn) is present at all times. We also detect coronal lines, indicative of a dense circumstellar medium. We constrain the progenitor wind velocity to  $\sim50$--60\,km\,s$^{-1}$ based on P-Cygni profiles, which is far slower than those present in Wolf-Rayet stars. This may suggest that the progenitor passed through a luminous blue variable phase, or that the wind is instead from a binary companion red supergiant star. 
At late times we see the $\sim2000$\,km\,s$^{-1}$ wide H$\alpha$ emission persisting at high luminosity ($\sim3\times10^{40}$\,erg\,s$^{-1}$) for at least 100 day, perhaps indicative of additional mass loss at high velocities that could have been ejected by a pulsational pair instability.
\end{abstract}

\keywords{supernovae: general --- supernovae: individual (SN~2017ens)}

\section{Introduction}
\label{sec:intro}

Type Ic supernovae (SNe) arise from the core collapse of a massive star that has lost its hydrogen and helium layers prior to exploding, through either strong stellar winds or interaction with a binary companion \citep[e.g.,][]{1997ARA&A..35..309F, 2017hsn..book..195G}. Their light curves are powered by the radioactive decay of $^{56}$Ni that is produced in the SN explosion.
Related to these events, but with luminosities up to 100 times higher, are the Type I superluminous SNe (SLSNe~I; see \citealt{2012Sci...337..927G, 2018ApJ...854..175I, 2018SSRv..214...59M} for reviews of observations and models). SLSNe exhibit spectral similarities to SNe~Ic \citep{2010ApJ...724L..16P}, but their luminosities are such that they cannot be powered solely by radioactive decay \citep{2011Natur.474..487Q}. The nature of the additional energy source remains unknown, with suggestions ranging from a central engine \citep{2010ApJ...717..245K,2010ApJ...719L.204W} to interaction with a massive H and He-free circumstellar medium \citep[CSM;][]{2011ApJ...729L...6C}.

Some SNe~Ib/Ic have been observed to develop relatively narrow ($\sim 500-1000$\,km\,s$^{-1}$) emission lines of hydrogen in their spectra; examples include SNe~Ib 2014C and 2004dk \citep{2015ApJ...815..120M, 2018arXiv180307051M} and SNe~Ic 2001em and 2017dio \citep{2017hsn..book..195G, 2018ApJ...854L..14K}. 
This has been interpreted as evidence that for at least some H-poor SNe, the fast ejecta are colliding with H-rich material relatively far from the star.
This late-time interaction has also been observed in some SLSNe~Ic which show H$\alpha$ emission at $+70$ to $+250$\,d after their peak brightness
\citep{2015ApJ...814..108Y,2017ApJ...848....6Y}. 

In this Letter we report on the discovery of an unusual SN with our GREAT ({\bf GR}OND-{\bf e}PESSTO-{\bf AT}LAS; \citealt{2008PASP..120..405G, 2015A&A...579A..40S, 2018PASP..130f4505T}) Survey. We introduce this program here, which is designed to rapidly identify hot, blue transients, with the specific goal of finding very young SLSNe in faint galaxies \citep{2017ATel10510....1C}. SN~2017ens (ATLAS17gqa) was discovered by the ATLAS survey on 2017 June 5 (UT dates are used herein), located at (J2000) $\alpha=12^h04^m09^s.37$, $\delta=-01^\circ55'52.2''$. 
Prompted by the high blackbody temperature of $21,000\pm3000$\,K that we measured with our GREAT data on 2017 June 8 \citep{2017ATel10478....1C}, we began an intensive spectroscopic and photometric follow-up campaign (Sec.\,\ref{sec:obs}). 

The adopted redshift of SN~2017ens, $z=0.1086$ (Sec.\,\ref{sec:analysis_result_coronal}), implies an absolute magnitude of $M_{g}=-21.1$ at peak, and thus a luminosity consistent with a SLSN \citep{2012Sci...337..927G}. 
In Sec.\,\ref{sec:analysis_result} we present the spectral evolution of SN~2017ens, which began to show $\sim2000$\,km\,s$^{-1}$ wide H$\alpha$ and H$\beta$ emission after $+163$\,d (phases are corrected for time dilation and are relative to the GROND $r$-band maximum on MJD = 57,924.011).
We compare the spectral properties of SN~2017ens to those of other SLSNe and broadlined SNe~Ic (SNe~Ic-BL), and also present the detections of rarely seen coronal lines. The bolometric light curve and modeling results are described in Sec.\,\ref{sec:LC_model}. 
Finally, in Sec.\,\ref{sec:dis} we discuss plausible scenarios that may explain the spectral evolution and luminosity of SN~2017ens.
We adopt a cosmology of $H_{0}=72~\mathrm{km~s^{-1}~Mpc^{-1}}$, $\Omega_\Lambda=0.73$, and $\Omega_{m}=0.27$. The foreground reddening toward SN~2017ens is $A_{V}=0.058$\,mag \citep{2011ApJ...737..103S}, and we assume that host-galaxy extinction is negligible because no Na~I~D absorption is visible in the SN spectrum.

\section{Observations}
\label{sec:obs}

Our photometric coverage of SN~2017ens 
spans the ultraviolet (UV) with the Ultraviolet and Optical Telescope (UVOT) on the Neil Gehrels {\it Swift} Observatory, optical wavelengths with GROND, ATLAS, LCO 1\,m\footnotemark[1], and Lulin Super Light Telescope (SLT)\footnotemark[2], and near-infrared (NIR) bands with GROND. 
We use standard procedures to reduce the data (\citealt{2008MNRAS.383..627P} for UVOT; \citealt{2008ApJ...685..376K} for GROND).
Ground-based optical photometry is calibrated against the Sloan Digital Sky Survey (SDSS). For ATLAS magnitudes we apply passband corrections using spectra \citep[prescription from][]{2018MNRAS.475.1046I}; for SLT data we use the conversion of R. Lupton\footnotemark[3]. The NIR magnitudes are calibrated against Two Micron All Sky Survey (2MASS) field stars. All data are reported in the AB system, and errors include the statistical and systematic uncertainties.
We do not have host-galaxy templates, but we estimate a $<15\%$ contribution from host light ($r>23$\,mag measured in pre-explosion Panoramic Survey
Telescope and Rapid Response System (PanSTARRS) images) to our SN photometry after $+150$\,d. 
Our photometric results are given in a machine-readable table and shown in Fig.\,\ref{fig:LC} (top panel).
\footnotetext[1]{https://lco.global/observatory/sites/}
\footnotetext[2]{http://www.lulin.ncu.edu.tw/slt76cm/slt\_introdution.htm}
\footnotetext[3]{http://classic.sdss.org/dr4/algorithms/sdssUBVRITransform.html}

We obtained a series of spectra of SN~2017ens, following the SN evolution from $+4$\,d to $+265$\,d (log of observations in Table\,\ref{tab:sn_spec}). 
Spectra are reduced in the standard fashion (ALFOSCGUI pipeline\footnotemark[4] for ALFOSC) or using custom-built pipelines PyWiFeS \citep{2014Ap&SS.349..617C} for WiFeS, LPipe\footnotemark[5] for LRIS, \citet{2015A&A...581A.125K} for X-Shooter, and \citet{2015A&A...579A..40S} for EFOSC2. 
Finally, we correct the spectral-flux calibration against $r$-band photometry. The resulting calibration error estimated by comparing to $g$-band photometry is generally $<0.10$\,mag, with the exception of the WiFeS (0.15\,mag) and Keck (0.25\,mag) spectra. (Those data were taken at very high airmass, making flux calibration difficult.)
All spectra will be available through WISeREP \citep{2012PASP..124..668Y}.
\footnotetext[4]{http://sngroup.oapd.inaf.it/foscgui.html}
\footnotetext[5]{http://www.astro.caltech.edu/$\sim$dperley/programs/lpipe.html}

\section{Analysis and results}
\label{sec:analysis_result}

\subsection{Light Curves and Comparison}
\label{sec:analysis_result_LC}

The discovery epoch of SN~2017ens with $M_{r}\approx-19.8$\,mag is at MJD = 57,909.3. ATLAS monitored the field daily for 23\,d before discovery. From a deep image taken 3\,d before discovery ($M_{r}\approx-18.7$\,mag), we constrain the explosion date of SN~2017ens to MJD $=57,907.8\pm1.5$; thus, the rest-frame rise time is $\sim15$\,d.

\begin{figure}[htbp]
  \centering
  \begin{subfigure}{0.50\textwidth}
    \includegraphics[width=\columnwidth]{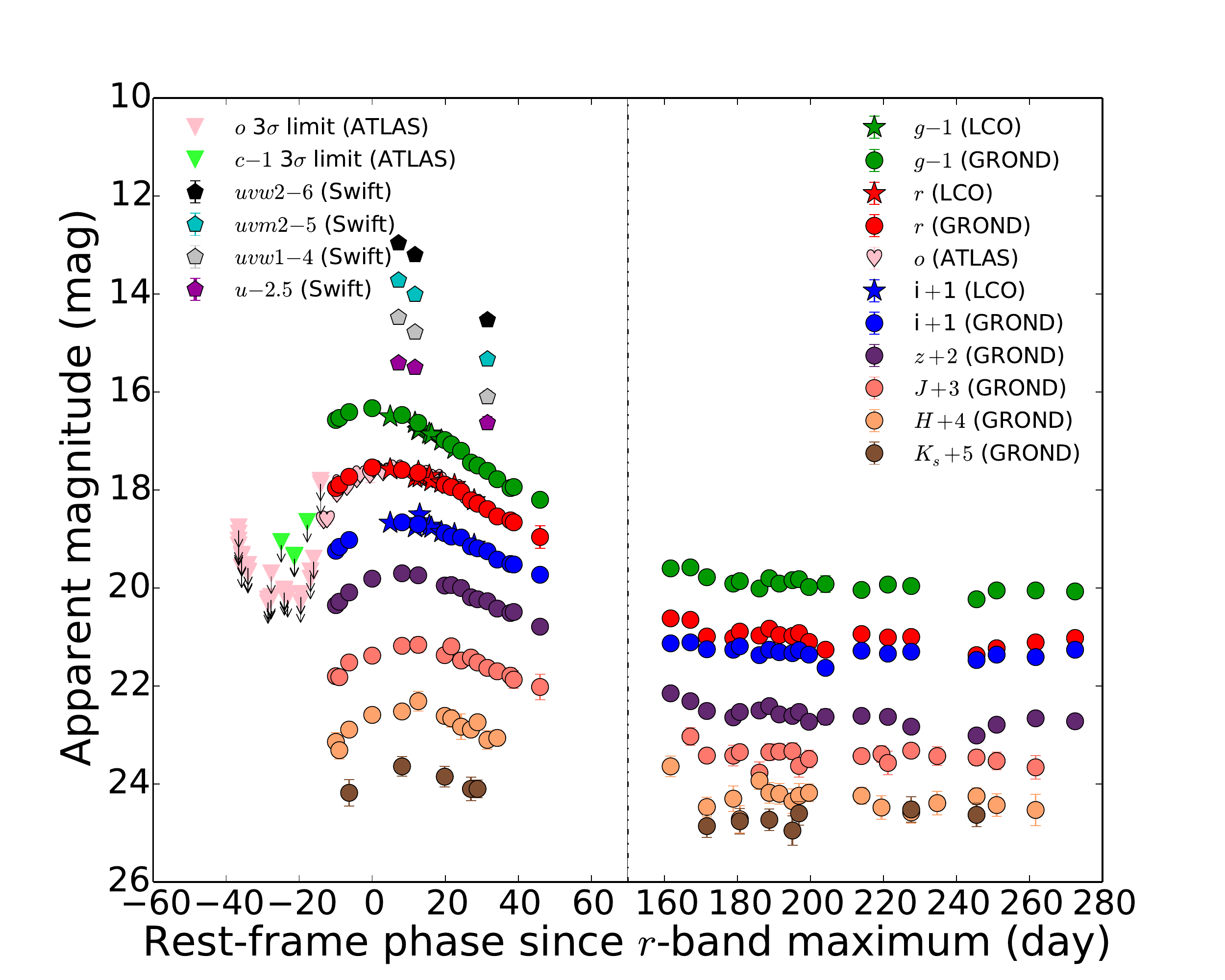}
  \end{subfigure}
   \begin{subfigure}{0.50\textwidth}
    \includegraphics[width=\columnwidth]{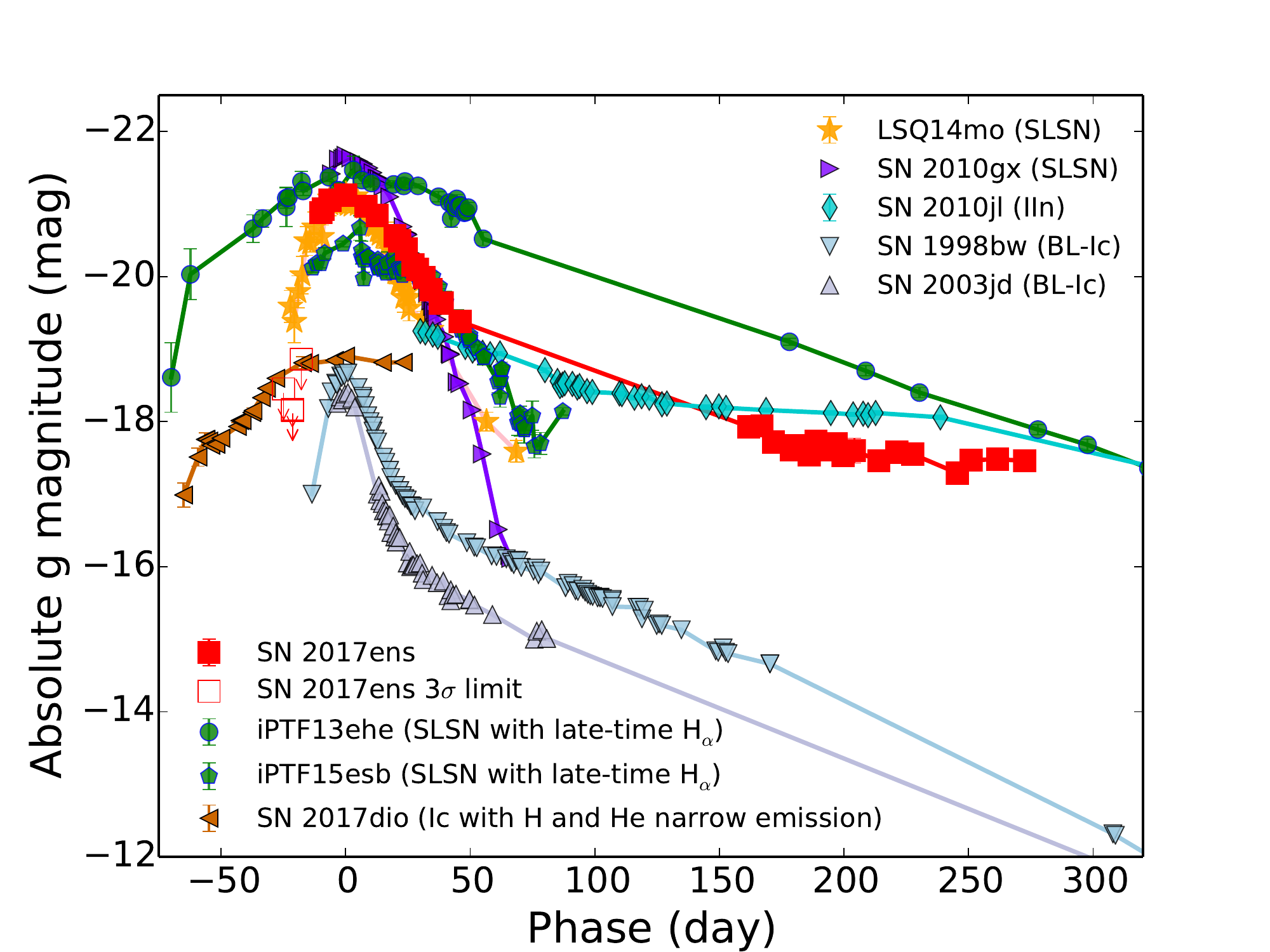}
  \end{subfigure}
  \begin{subfigure}{0.50\textwidth}
    \includegraphics[width=\columnwidth]{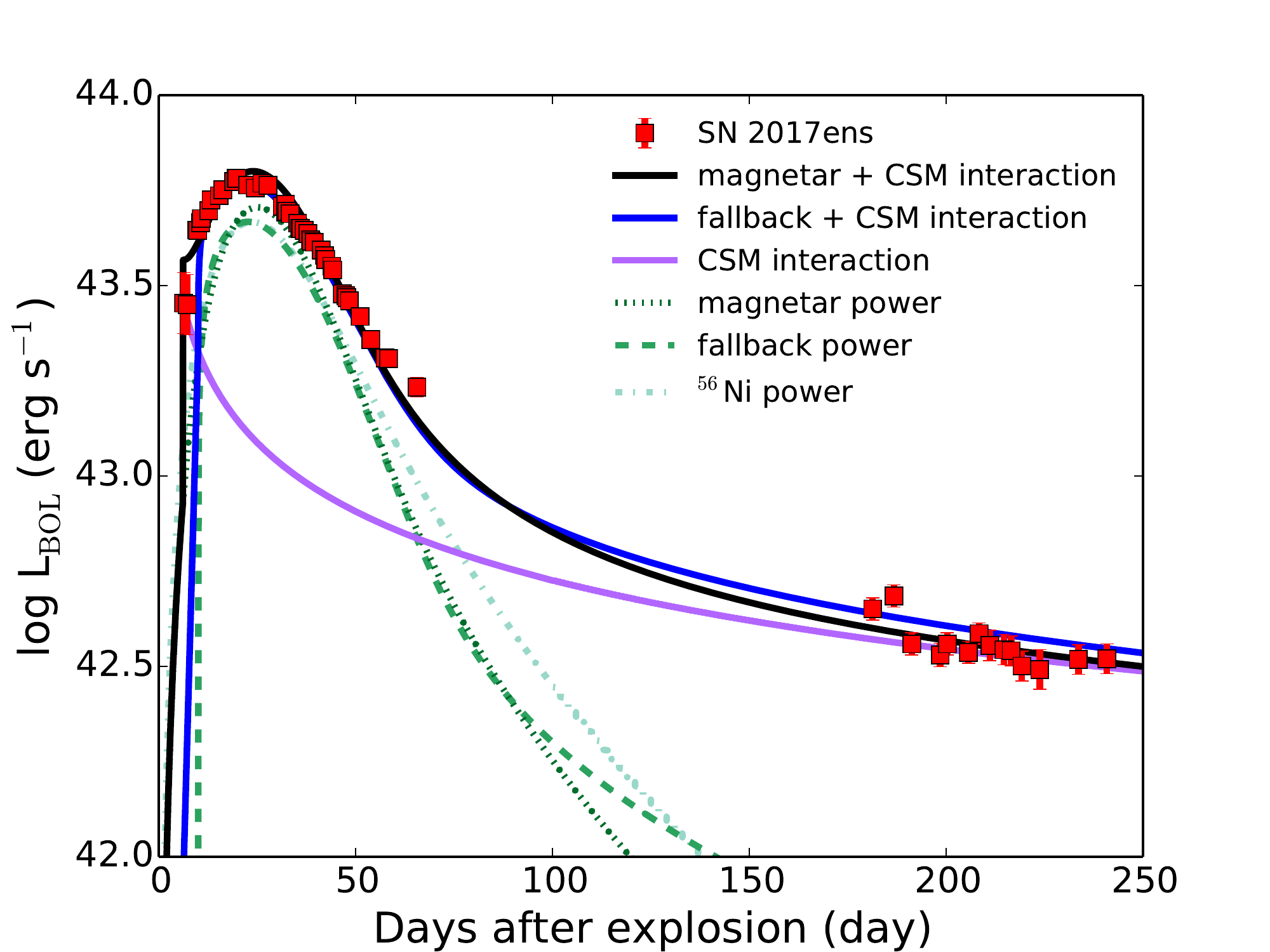}
  \end{subfigure}
  \caption{{\it Top panel:} multiband light curves of SN~2017ens. Note the discontinuous abscissa. {\it Middle panel:} light-curve comparison in absolute $g$ band with the SNe chosen for spectroscopic comparison. {\it Bottom panel:} bolometric light curve of SN~2017ens and model fitting.}
  \label{fig:LC}
\end{figure}

Fig.\,\ref{fig:LC} (middle panel) shows the absolute $g$-band light curve, which we compare to SLSNe, SNe~IIn, and SNe~Ic-BL selected based on the photometric properties and spectral evolution (see Sec.\,\ref{sec:analysis_result_spec}) of SN~2017ens.
At peak, SN~2017ens is $\sim10$ times more luminous than the SNe~Ic-BL~1998bw \citep{2001ApJ...555..900P}, 2003jd \citep{2008MNRAS.383.1485V}, and SN~Ic~2017dio \citep{2018ApJ...854L..14K}, which shows narrow H and He emission in its spectra. 
The early-phase light-curve evolution of SN~2017ens is similar to that of rapidly evolving SLSNe such as LSQ14mo \citep{2017A&A...602A...9C, 2015ApJ...815L..10L} and SN~2010gx \citep{2010ApJ...724L..16P}. SN~2017ens shows no sign of undulations in its light curves, as are often observed in slowly evolving SLSNe as well as SLSNe that exhibit late-time H$\alpha$ such as iPTF13ehe \citep{2015ApJ...814..108Y} and iPTF15esb \citep{2017ApJ...848....6Y}. 
At late times, the light curves of SN~2017ens remain approximately constant, indicating that strong interaction dominates, as in SN~IIn~2010jl before +300\,d \citep{2014ApJ...797..118F}.

\subsection{Spectroscopic Evolution and Comparison}
\label{sec:analysis_result_spec}

\begin{figure*}
    \centering
       \includegraphics[width=\textwidth]{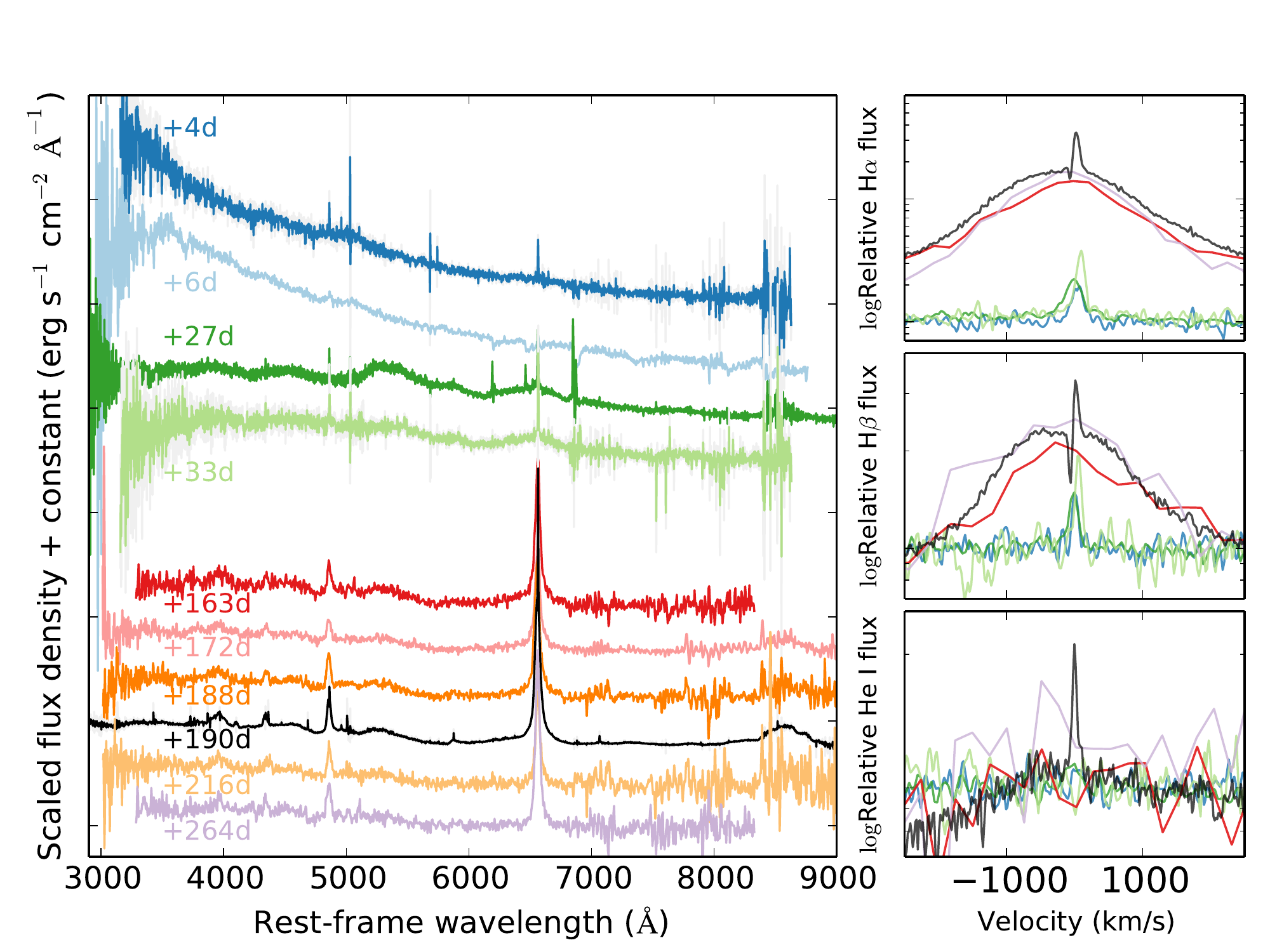}
\caption{Spectroscopic evolution of SN~2017ens. The right panels show the velocity of the H$\alpha$, H$\beta$, and \ion{He}{1} $\lambda 5876$ lines at selected epochs. Each phase is shown with the same color as in the main panel.}
\label{fig:spec_evolution}
\end{figure*}

We show the spectral evolution of SN~2017ens in Fig.\,\ref{fig:spec_evolution}.
Around maximum light the spectra are blue and featureless. In the first spectrum taken at $+4$\,d after peak, we detect narrow H$\alpha$ and H$\beta$ emission lines (barely resolved width of $\sim100$\,\kms). Fitting the dereddened spectra with a blackbody gives a temperature of $T_{\rm BB}\geq10,300$\,K, consistent with our estimate from the GROND analysis ($\geq11,500$\,K).  
At $\sim1$ month after peak, some broad features emerge, similar to those seen in SNe~Ic-BL after peak brightness \citep[e.g.,][]{2001ApJ...555..900P}.
Apart from narrow H$\alpha$ and H$\beta$, we detect a narrow \ion{He}{1} $\lambda5876$ emission line.
The commonly observed [\ion{O}{2}], [\ion{O}{3}], and [\ion{N}{2}] host-galaxy emission lines are absent, suggesting that the observed Balmer lines originate from the transient itself, not the underlying host \citep{2017ATel10587....1P}. We also check the WiFeS datacubes and see no [\ion{O}{3}] emission at the SN position.

At late times ($>160$\,d) after the SN emerged from solar conjunction, our data reveal dramatic evolution, with the spectra more resembling those of SNe~IIn. The spectra are still blue, but now dominated by prominent, $\sim2000$\,km\,s$^{-1}$ wide Balmer emission lines, indicative of much stronger interaction with H-rich CSM. The luminosity and the velocity of the $\sim2000$\,km\,s$^{-1}$ H$\alpha$ line does not vary significantly between $+163$ and $+264$\,d, staying at $\sim3\times 10^{40}$\,erg\,s$^{-1}$.

\begin{figure*}
    \centering
    \begin{subfigure}{1\linewidth}        
        \centering
        \includegraphics[width=0.65\linewidth]{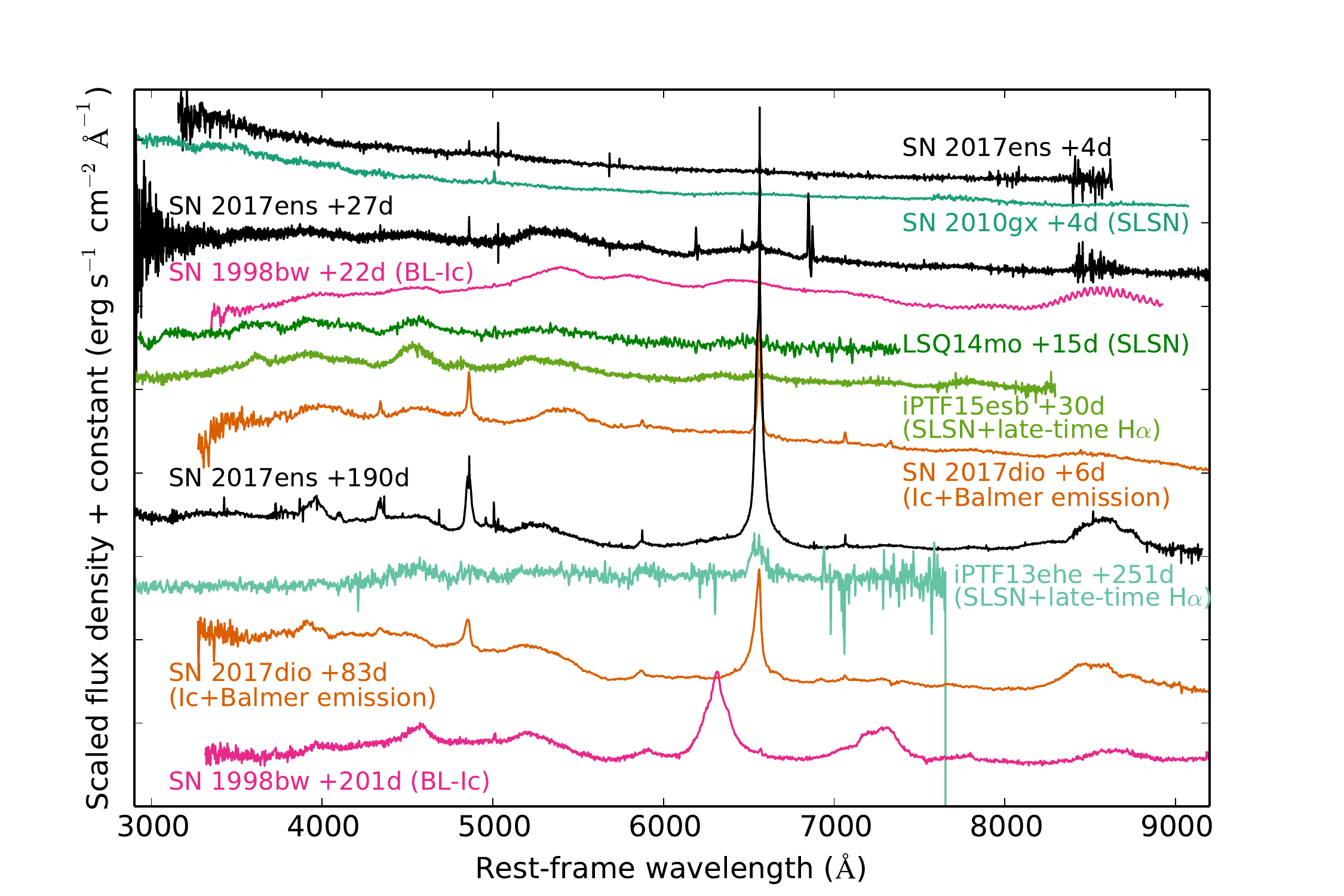}
    \end{subfigure}
    \begin{subfigure}{1\linewidth}        
        \centering
        \includegraphics[width=0.65\linewidth]{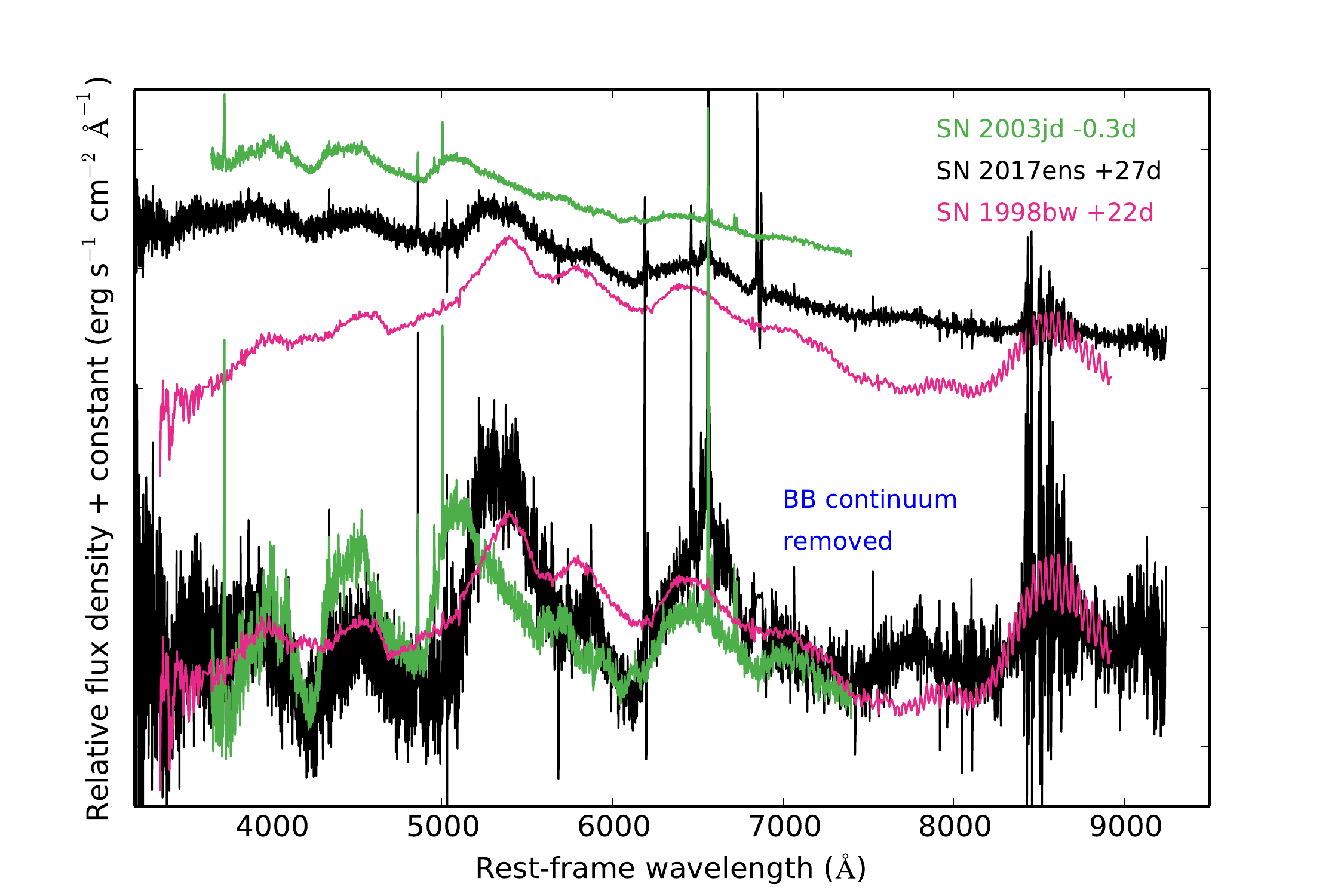}
    \end{subfigure}
    \begin{subfigure}{1\linewidth}        
        \centering
        \includegraphics[width=0.65\linewidth]{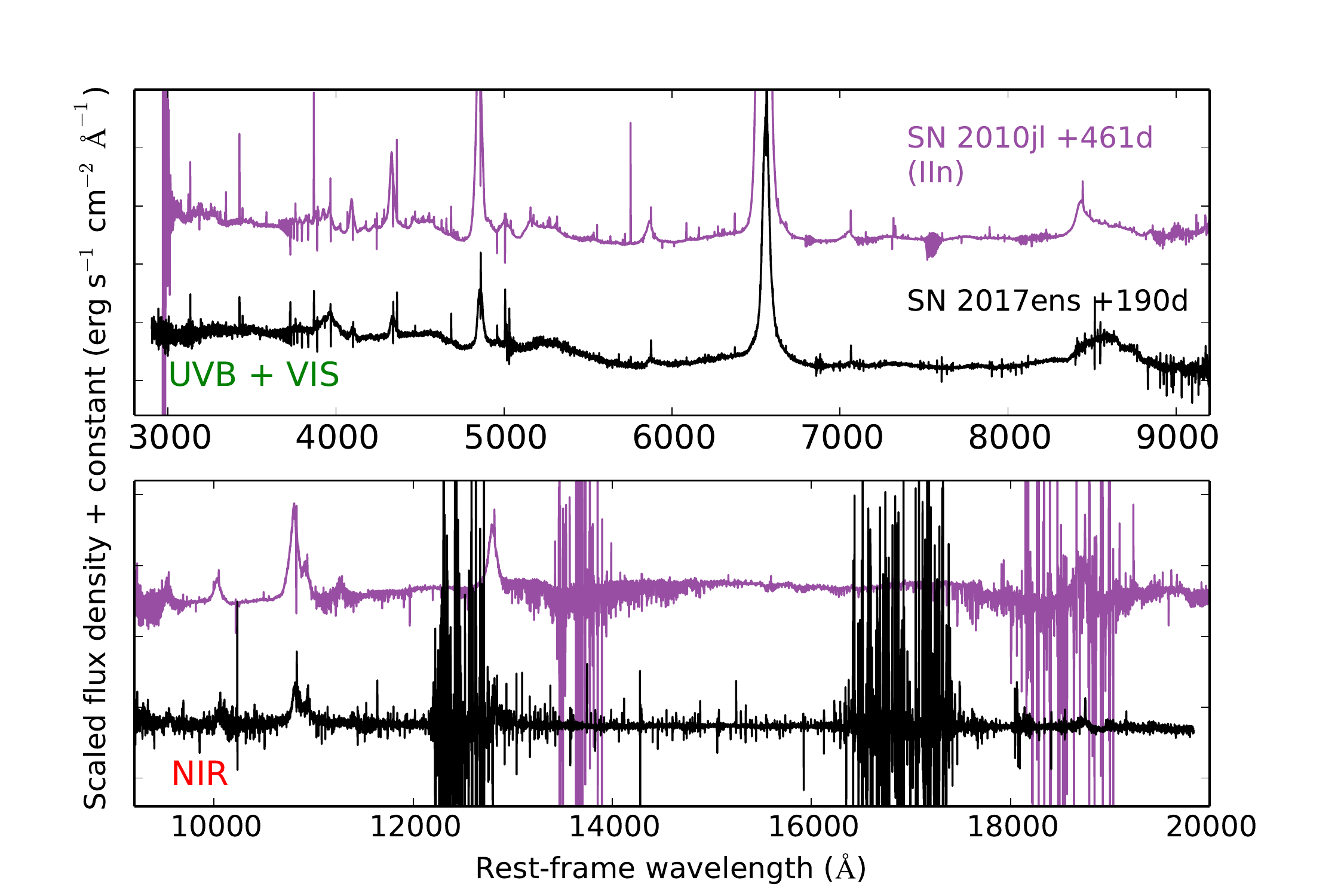}
    \end{subfigure}
    \caption{{\it Top panel:} comparison between SN~2017ens and other SNe at three selected epochs. 
 {\it Middle panel:} comparison between SN~2017ens and the Type Ic-BL SNe~1998bw and 2003jd. The lower spectra have their continuum removed assuming a blackbody.
{\it Bottom panel:} comparison between SN~2017ens and the Type IIn SN~2010jl at optical and NIR wavelengths.}
    \label{fig:spec_comp}
\end{figure*}

The spectral evolution of SN~2017ens is unique, sharing features with several distinct SN subclasses (Fig.\,\ref{fig:spec_comp} top panel). 
In the earliest phases, the blue and featureless spectra share a similarity with young core-collapse SN spectra. 
We do not see the \ion{O}{2} absorption features commonly associated with SLSNe. However, we may have missed them in SN~2017ens. For example, SLSN~2010gx \citep{2010ApJ...724L..16P} displayed \ion{O}{2} absorption before it peaked and then became blue and featureless. 

As the spectra evolve, SN~2017ens is not well matched to other SLSNe such as LSQ14mo \citep{2017A&A...602A...9C} and iPTF15esb \citep{2017ApJ...848....6Y}. Rather, it appears to be more similar to SNe~Ic-BL. The classification tool GELATO \citep{2008A&A...488..383H} applied to the SN~2017ens $+27$\,d spectrum returns the closest similarity with SN~1998bw at $+22$\,d \citep{2001ApJ...555..900P} and SN~2003jd at $-0.3$\,d \citep{2014AJ....147...99M}. 
These two SNe~Ic-BL still provide a good match to SN~2017ens when we remove the continua assuming a blackbody (Fig.\,\ref{fig:spec_comp}, middle panel).
SN~2017ens has a somewhat bluer continuum, perhaps due to CSM interaction, as was the case for SN~2017dio at +6\,d \citep{2018ApJ...854L..14K}. 
The origin of the broad feature around 6530\,\AA\ is uncertain; it could be attributed to a blend of Si and Fe/Co lines, H$\alpha$ associated with interaction, or the \ion{C}{2}  $\lambda$6580 line sometimes seen in SLSNe \citep[e.g., SN~2018bsz;][]{2018arXiv180610609A}.

During the late-time strongly interacting phase, the overall spectral features of SN~2017ens are well matched with those of SN~2017dio at +83\,d. Both SNe exhibit a blue pseudocontinuum (below $\sim5000$\,\AA) that is more significant than in iPTF13ehe at +251\,d \citep{2015ApJ...814..108Y}; it is likely produced by \ion{Fe}{2} lines \citep{2009ApJ...695.1334S}.

\subsection{Nebular and Coronal Lines}
\label{sec:analysis_result_coronal}

\begin{figure*}
\centering
\begin{subfigure}[b]{0.6\textwidth}
   \includegraphics[width=1\linewidth]{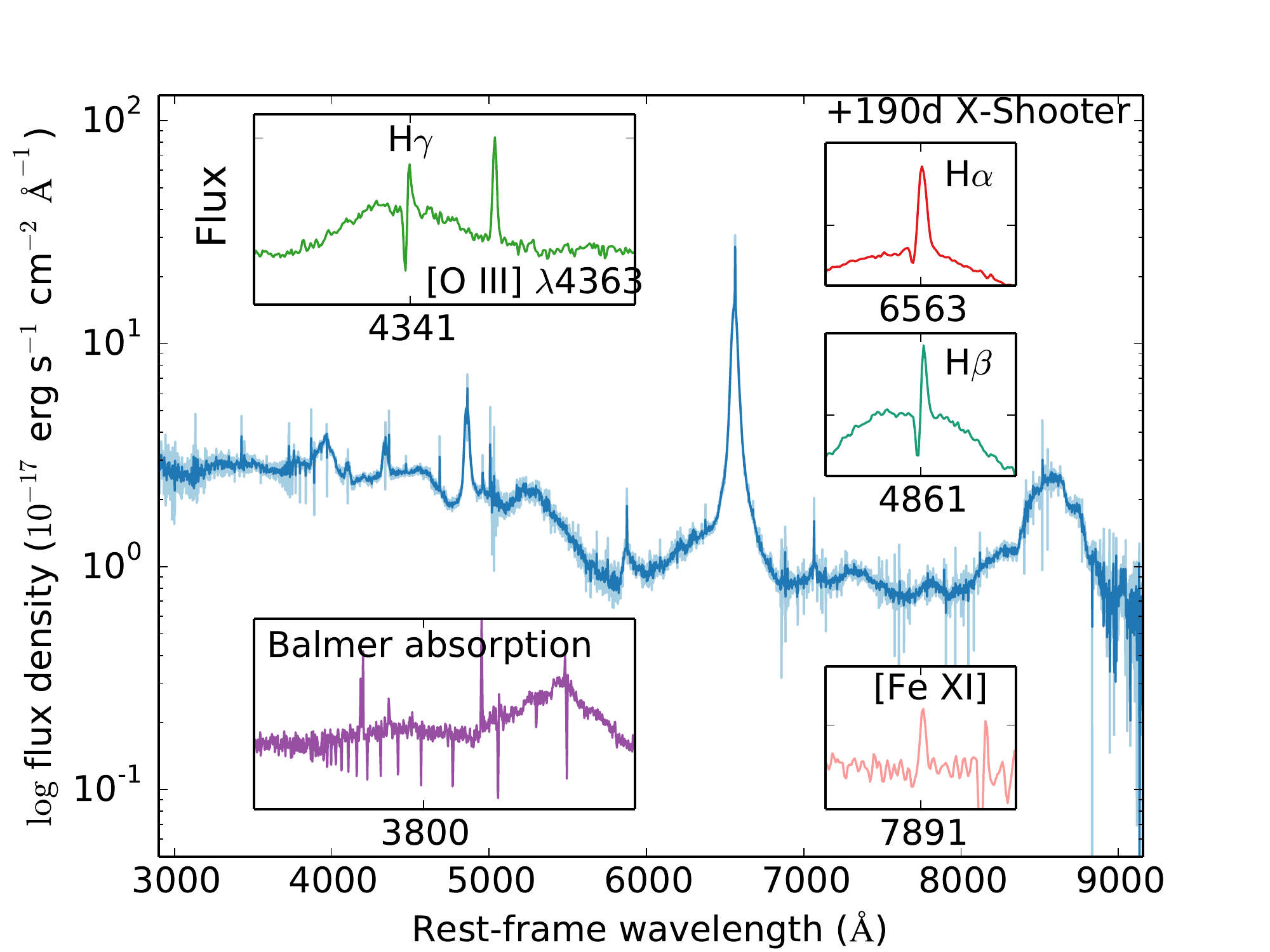}
\end{subfigure}
\begin{subfigure}[b]{0.6\textwidth}
   \includegraphics[width=1\linewidth]{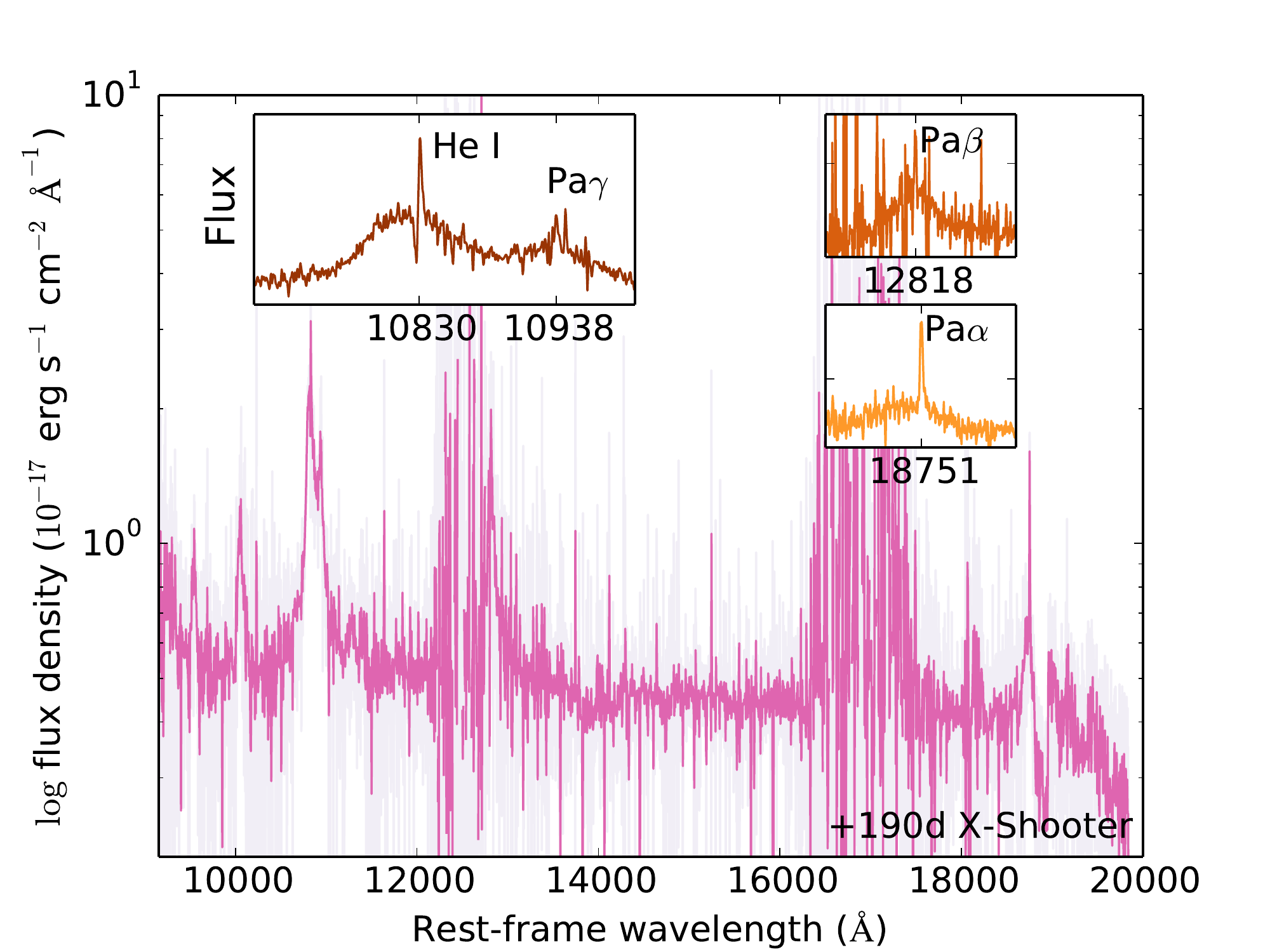}
\end{subfigure}
\begin{subfigure}[b]{0.45\textwidth}
   \includegraphics[width=1\linewidth]{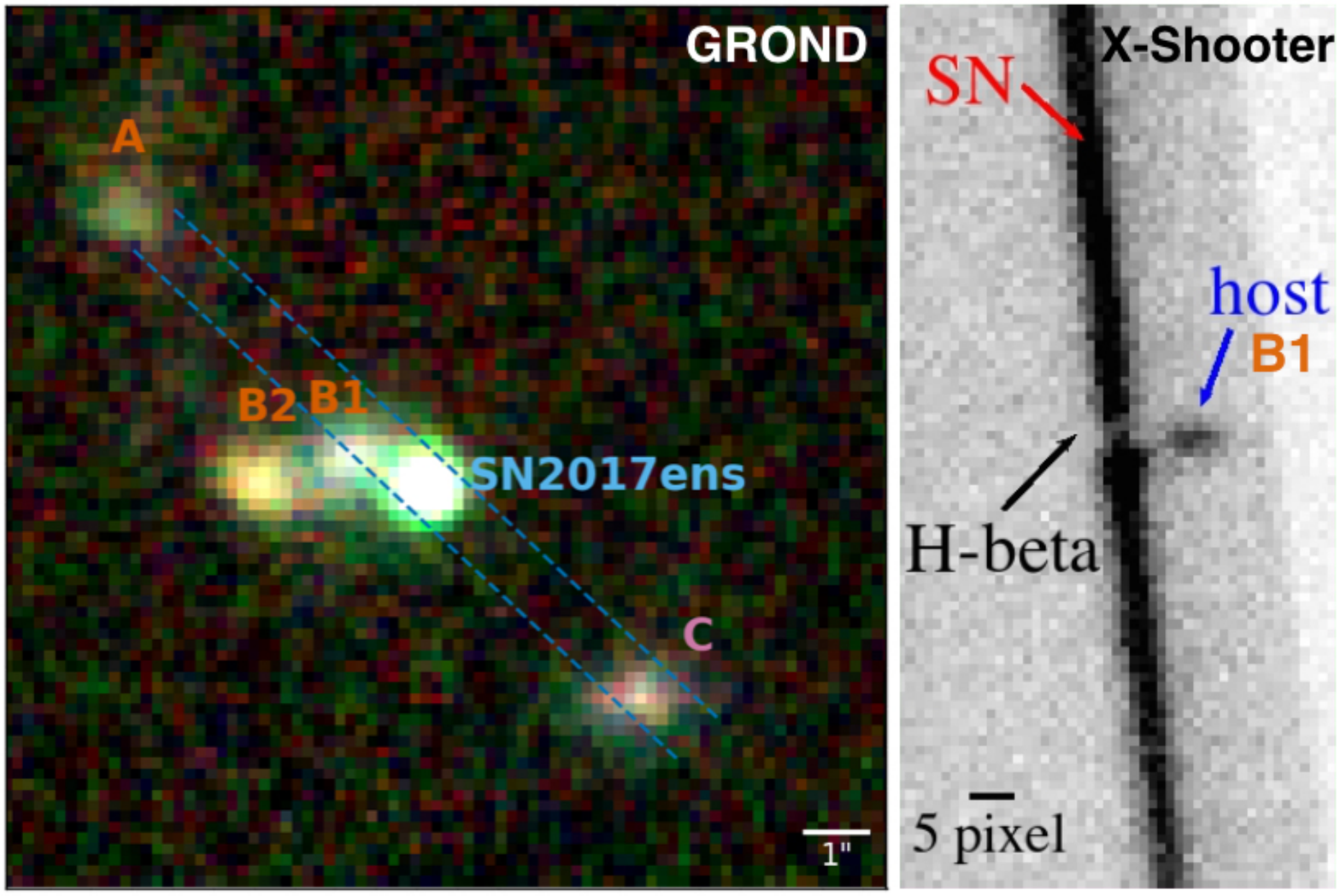}
\end{subfigure}
\caption{VLT/X-Shooter spectra of SN~2017ens at $+190$\,d. {\it Top panel:} UVB and optical (VIS) parts. {\it Middle panel:} NIR part. The main panels use a log scale in order to present details in the continuum (rebinned to 2\,\AA\,pixel$^{-1}$ (UVB+VIS) and 5\,\AA\,pixel$^{-1}$ (NIR)); inset panels use a linear scale without binning. {\it Bottom panel left:} color-combined image from GROND $r'i'z'$ bands. SN~2017ens is associated with host-galaxy B1+B2, and there is a possible tidal tail connecting to the nearby object A (redshift unknown). Source C is a background galaxy at $z=0.30$. The X-Shooter slit position is indicated with two dashed lines. {\it Bottom panel right:} 
the H$\beta$ position in the two-dimensional raw image.}
\label{fig:spec_vlt} 
\end{figure*}

The VLT/X-Shooter spectra around $+190$\,d (Fig.\,\ref{fig:spec_vlt}) provide higher resolution and wider wavelength coverage than our other spectra, enabling us to detect many narrow emission lines. 
Interestingly, we find that the flux ratio of the nebular [\ion{O}{3}] $\lambda\lambda4959$, 5007 and auroral [\ion{O}{3}] $\lambda4363$ lines is 0.45, consistent with coronal lines that may arise from X-ray photoionization \citep{2002ApJ...572..350F} of dense gas (see \citealt{1984ApJ...285..458F}, their Fig.\,11).
Therefore, we conclude that the [\ion{O}{3}] $\lambda4363$ line comes from the SN, and we use it to constrain the redshift of SN~2017ens to $z=0.1086$, consistent with the average of the [\ion{O}{2}] $\lambda3727$ and [\ion{O}{3}] $\lambda\lambda4959$, 5007 lines.

These narrow coronal lines have been seen in only a handful of SNe~IIn and the transitional object SN~2011hw \citep{2015MNRAS.449.1921P}.
The ratio [\ion{O}{3}]$\lambda$4363/[\ion{O}{3}]$\lambda\lambda$4959, 5007 for SN~2017ens is similar to that seen in SN~2005ip at +173\,d \citep{2009ApJ...695.1334S}, SN~2006jd at $+1542$\,d \citep{2012ApJ...756..173S}, and SN~2010jl at $+461$\,d and $+573$\,d \citep{2014ApJ...797..118F}.
Other coronal lines detected in SN~2017ens are similar to those seen in SN~2010jl (Fig.\,\ref{fig:spec_comp}, bottom panel):
[\ion{Fe}{10}] $\lambda6374.5$ is strong, as are [\ion{Fe}{11}]
$\lambda7891.8$, [\ion{Ne}{5}] $\lambda\lambda3345.8$, 3425.9, [\ion{Ca}{5}] $\lambda6086.8$, and [\ion{Ar}{10}] $\lambda5533.2$. 
The presence of these lines is indicative of a highly ionized and dense CSM, although we do not detect the highest-ionization coronal lines such as [\ion{Fe}{14}] $\lambda5302.9$ and [\ion{Ar}{14}] $\lambda4412.3$, which were seen in SN~2005ip. 

The flux ratio of the [\ion{O}{3}] $\lambda4363$ to $\lambda5007$ lines is a function of the CSM density and temperature. Following \citet[][their Fig.\,26]{2014ApJ...797..118F}, we use our measured flux ratio, log($\lambda4363/\lambda5007)=-0.22$, to constrain the CSM electron density to lie between $10^{6}$ and $10^{8}$\,cm$^{-3}$ for $T_e =50,000$ to 10,000\,K. This density range is consistent with that observed for SN~2010jl. 

From our mid-resolution X-Shooter data, we resolve narrow P-Cygni profiles on top of the $\sim2000$\,km\,s$^{-1}$ wide Balmer and Paschen lines. 
We measure the blueshifted wavelength from the absorption component of the H$\gamma$, H$\beta$, and H$\alpha$ P-Cygni profiles, which suggests that the unshocked CSM has a low velocity of $\sim50$\,km\,s$^{-1}$. 
A similar velocity of $\sim60$\,km\,s$^{-1}$ is obtained from the P-Cygni profile of the \ion{He}{1} $\lambda10,830$ line.
Moreover, we measure the full width at half-maximum intensity (FWHM) of the wide components, such as H$\alpha$ ($2500\pm700$\,km\,s$^{-1}$), H$\beta$ ($2300\pm400$\,km\,s$^{-1}$), Pa$\gamma$ ($2000\pm200$\,km\,s$^{-1}$), and \ion{He}{1} $\lambda10,830$ ($2200\pm200$\,km\,s$^{-1}$). 
We also detect narrow absorption lines from the Balmer series (no clear emission), spanning H$\epsilon$ to H33 (3659\,\AA). 

In addition, we see emission from the \ion{H}{2} region close to the host-galaxy center (see Fig.\,\ref{fig:spec_vlt}, marked B1), as part of a faint galaxy (SDSS J120409.47--015552.4) with 
$g = 21.92\pm0.24$\,mag ($M_{g}\approx-16.5$\,mag). 
These lines have a slightly different redshift ($z=0.1084$) than SN~2017ens.
In particular, the (noisy) detection of the weak auroral [\ion{O}{3}] $\lambda4363$ line indicates a low host-galaxy metallicity of $\sim0.04$--$0.4\,Z_{\odot}$ using the direct ${\rm T_{e}}$-based method. If we instead use the empirical N2 metallicity diagnostic \citep{2004MNRAS.348L..59P}, we measure $Z=0.3\pm0.2\,Z_{\odot}$.

\section{Bolometric Light Curve and Model Fitting} 
\label{sec:LC_model}

Using all of our available UV-through-NIR photometry, we built a pseudobolometric light curve for SN~2017ens using the prescription from \citet{2018MNRAS.475.1046I}. The results are very similar to those derived when using a blackbody fit, as expected as our photometry covers a large wavelength range.
From a polynomial fit to the bolometric data we obtain $L_{\rm bol}=(5.86\pm0.20)\times10^{43}$\,erg\,s$^{-1}$ at peak and an integrated energy of $(3.53\pm1.42)\times10^{50}$\,erg. 

To fit our bolometric light curve, we used a two-component model consisting of a central heating and an interaction component. First, the centrally heated component uses the standard Arnett method \citep{1982ApJ...253..785A,2013ApJ...770..128I}. We tested three possible
central power sources: the nuclear decay of $^{56}$Ni, the spindown of a magnetar \citep{2010ApJ...717..245K}, and fallback accretion \citep{2013ApJ...772...30D, 2018arXiv180600090M}. The $^{56}$Ni decay and the magnetar spindown light curves are obtained as by \citet{2013ApJ...770..128I}, but the magnetar model takes the gamma-ray opacity from the magnetar into account as by \citet{2015MNRAS.452.1567C}. The fallback accretion power is obtained by assuming a central energy input of $L_{\mathrm{fallback},1}(t/1\,\mathrm{s})^{-5/3}$, where $L_{\mathrm{fallback},1}$ is a constant \citep{2013ApJ...772...30D}. Second, for the interaction component, we adopted a steady-wind CSM, and the input luminosity from this component goes as $L_{\mathrm{int},1}(t/1\,\mathrm{s})^{-3/5}$, where the outer SN density structure is proportional to $r^{-7}$ \citep{2013MNRAS.435.1520M}. The inner SN density structure is assumed to be constant.

We first used the interaction component to fit the bolometric light curve 150\,d after explosion, assuming that interaction is the dominant light source at this time. We then derived the contribution required from a central power source at early times to provide a good light-curve match. Given that the spectra of SN~2017ens and SN~1998bw are similar (Fig.\,\ref{fig:spec_comp}),
we used the relation $(E_\mathrm{ej}/10^{51}\,\mathrm{erg})/(M_\mathrm{ej}/M_\odot)\approx3$ found for SN~1998bw \citep{2001ApJ...550..991N} to break the degeneracy between $E_\mathrm{ej}$ and $M_\mathrm{ej}$.

Fig.\,\ref{fig:LC} (bottom panel) shows the results of our fits. In all cases, the CSM interaction model that we used has $L_{\mathrm{int},1}=7.7\times10^{46}\,\mathrm{erg\,s^{-1}}$. 
The inner edge of the CSM is set at $1.2\times10^{15}\,\mathrm{cm}$ to match the early light-curve rise in the model, but this constraint is not strong.
We find that all three centrally heated models provide reasonable fits to the bolometric light curve. They all have $E_\mathrm{ej}=1.5\times10^{52}\,\mathrm{erg}$ and an ejecta mass of $5\,M_\odot$. 
However, the $^{56}$Ni-powered light curve requires a very high $^{56}$Ni mass of $3.5\,M_\odot$. This is close to the ejecta mass, and we therefore find the $^{56}$Ni-powered model to be unlikely. Alternatively, a magnetar central engine with an initial spin of 3.8\,ms and a magnetic field of $8\times10^{13}\,\mathrm{G}$, and fallback accretion with $L_{\mathrm{fallback},1}=6\times10^{53}\,\mathrm{erg\,s^{-1}}$, provide good qualitative fits to the light curve. It is of course possible that the entire light curve is driven by different degrees of interaction. The contribution of the interaction component at early times (0--70\,d after explosion) is $\sim20$\%, while it is $\gtrsim90$\% at late times (200\,d).

Assuming the above best-fit results and a kinetic energy to radiation conversion efficiency at the shock of 0.1 \citep{2013MNRAS.435.1520M}, we estimate the mass-loss rate of the progenitor to be $5\times10^{-4}\,M_\odot\,\mathrm{yr^{-1}}$, with a constant wind velocity of $50\,\mathrm{km\,s^{-1}}$. The CSM density estimate is similar to those of SNe~IIn showing similar coronal lines \citep{2013A&A...555A..10T}.

\section{Discussion}
\label{sec:dis}

One important clue to interpreting the possible powering mechanisms behind SN~2017ens is that we measured the H-rich material to have a velocity of $\sim50$--60\,\kms\ from the blueshifted absorption of the narrow P-Cygni profiles. This wind velocity is far slower than those present in Wolf-Rayet star winds. 
If this wind is from the progenitor, it could come from a massive H-rich progenitor (such as a luminous blue variable) that explosively ejected its H envelope shortly before the SN explosion. Alternatively, this wind could come from a pulsational pair-instability SN with a
slow and long-term stable wind \citep{2017ApJ...836..244W}.

It is also possible that SN~2017ens exploded as a SN~Ic-BL inside a patchy, H-rich CSM from a binary companion; the expanding ejecta interact with the bulk of the CSM at later times, as has been suggested for SN~2017dio \citep{2018ApJ...854L..14K}. Alternatively, as proposed for ASASSN-15no \citep{2018MNRAS.476..261B}, a dense inner CSM may have hidden the SN features at early times, before they become briefly visible as the CSM was swept up by the ejecta. At late times they could have again been masked by an increasingly strong interaction component. A special CSM geometry (e.g., doughnut shape) is also probable, and we see the SN~Ic-BL along a certain viewing angle.

In the case of a binary companion,
the wind of $\sim50$--60\,\kms\ and mass-loss rate of $5\times10^{-4}\,M_\odot\,\mathrm{yr^{-1}}$ are consistent with a red supergiant \citep{2017MNRAS.465..403G}, albeit at the more extreme end, which can be explained by the companion having gained mass from the SN progenitor during an earlier accretion phase.
If so, this may suggest that the progenitor of SN~2017ens lost its H and He layers through interaction with a binary companion. 

We must also consider the apparent $\sim2000$\,\kms\ material given its high luminosity. If this is associated with mass loss from the progenitor, and the line width is not from electron scattering as seen in many SNe~IIn, then the material is moving much faster than the winds of H-rich stars (or the CSM of SNe~IIn). It is difficult to imagine how this could be produced by anything other than a sudden ejection of the H envelope, shortly before the SN explosion. In fact, the luminosity of the $\sim2000$\,km\,s$^{-1}$ wide component of H$\alpha$ is comparable to that seen in SN~1995N \citep{2002ApJ...572..350F} ($\sim2.3\times10^{40}$\,erg\,s$^{-1}$), and it may be too large to be coming solely from swept-up material. 
A pulsational pair-instability explosion is at least qualitatively consistent with an outburst that can unbind the H envelope shortly before an SN explosion. This scenario is also consistent with the measured low-metallicity environment.

The unique spectroscopic evolution of SN~2017ens together with its high luminosity poses challenges to all currently known SN scenarios. While detailed modeling can help elucidate the nature of this transient, ongoing surveys for SLSNe such as GREAT will find more such peculiar transients. With a larger sample and high-cadence follow-up spectroscopy, we will be able to further understand the nature of SN 2017ens-like objects and the role of interaction in SLSNe.

\acknowledgments
T.W.C. acknowledges Thomas~Kr\"uhler for the X-Shooter data reduction, Lin Yan and Claes Fransson for providing comparison spectra, Jason Spyromilio for useful discussions, Chien-Hsiu~Lee and You-Hua~Chu for coordinating observational resources, and funding from the Alexander von Humboldt Foundation. 
M.F. acknowledges the support of a Royal Society -- Science Foundation Ireland University Research Fellowship. P.S. acknowledges support through the Sofia Kovalevskaja Award (Alexander von Humboldt Foundation). 
A.V.F. is grateful for the support of the TABASGO Foundation, the Christopher R. Redlich fund, and the Miller Institute for Basic Research in Science (U.C. Berkeley).
A.J.R. and I.R.S. are supported by the Australian Research Council through grants FT170100243 and FT160100028, respectively. F.T. and J.S. acknowledge support from the KAW Foundation. 
S.J.S. acknowledges funding from the European Research Council Grant agreement \#291222 and STFC grant ST/P000312/1. 
M.G. is supported by Polish National Science Centre grant OPUS 2015/17/B/ST9/03167. K.M. acknowledges support from the UK STFC through an Ernest Rutherford Fellowship and from a Horizon 2020 ERC Starting Grant (\#758638). L.G. was supported in part by US NSF grant AST-1311862. C.P.G. acknowledges support from EU/FP7-ERC grant \#615929. Z.Y.L., C.C.N., and P.C.Y. are grateful for funding from MoST (Taiwan) under grants 105-2112-M-008-002-MY3, 104-2923-M-008-004-MY5, and 106-2112-M-008-007. A.P. and S.B. are partially supported by PRIN-INAF 2017 ``Toward the SKA and CTA era: discovery, localization, and physics of transient sources'' (P.I.: Giroletti). A.G.-Y. is supported by the EU via ERC grant No. 725161, the Quantum Universe I-Core program, the ISF, the BSF Transformative program, and a Kimmel award.

Part of the funding for GROND was generously granted from the Leibniz Prize to Prof. G. Hasinger (DFG grant HA 1850/28-1). Partly observations made with the Nordic Optical Telescope using ALFOSC. This publication has made use of data collected at Lulin Observatory, partly supported by MoST grant 105-2112-M-008-024-MY3. 
Some of the data
presented herein were obtained at the W. M. Keck Observatory, which is
operated as a scientific partnership among the California Institute of
Technology, the University of California, and NASA; the observatory was made possible by
the generous financial support of the W. M. Keck Foundation.

\begin{sidewaystable}
 \begin{minipage}{190mm}
 \centering
  \caption{Log of spectroscopic observations of SN~2017ens and its host galaxy.} 
\label{tab:sn_spec}
\begin{tabular}[t]{llllllllll}
\hline
UT Date & MJD & Phase & Telescope & Instrument & Grating/Grism & Exp. Time   & Slit  & Resolution  & Range   \\
  &   &  (day) &   &   & /Arm &   (s) &  (\arcsec) &  (\AA) &   (\AA) \\
\hline
2017 Jun. 24 & 57928.392&	3.95 & ANU 2.3\,m  & WiFeS$^a$ & B3000/R3000 & 1200/1200 & IFU & 1.6/2.5 & 3500--5700/5400--9500 \\
2017 Jun. 26 & 57930.356 &	5.72 & NOT & ALFOSC &  Gr\#4 & 1800 & 1.0 & 16 & 3300--9700\\
2017 Jul. 20 & 57954.264 &	27.29 & Keck I & LRIS & B600/R400 & 1125 & 0.7 & 5/6 & 3200--10,000\\
2017 Jul. 26 & 57960.356 &32.78 & ANU 2.3\,m  & WiFeS$^a$ & B3000/R3000 & 1200/1200 & IFU & 1.6/2.5 & 3500--5700/5400--9500 \\
2017 Dec. 17 & 58104.325	& 162.65 & NTT & EFOSC2 &  Gr\#13 & 2700 & 1.0 & 18.2 & 3700-9200 \\
2017 Dec. 27 & 58114.301 &	171.65 & NTT & EFOSC2 & 
Gr\#11/Gr\#16 & 2700/2700  & 1.0/1.0 &  13.8/13.4 &  3400--7400/6000--9900\\
2018 Jan. 14	& 58132.275 & 187.86	& VLT  & X-Shooter & UVB/VIS/NIR & 3600/3400/3680 & 0.9/0.9/1.0 & 1/1.1/3.3 & 3000--5560/5450--10,200/10,000--20,600\\
2018 Jan. 14 & 58132.312	& 187.90 & NTT & EFOSC2 & 
Gr\#11 & 3600  & 1.0 &  13.8 &  3400--7400\\
2018 Jan. 15	& 58133.263& 188.75	& VLT  & X-Shooter & UVB/VIS/NIR & 7200/6800/7360 & 0.9/0.9/1.0 & 1/1.1/3.3 & 3000--5560/5450--10,200/10,000--20,600\\
2018 Jan. 15 & 58133.273 &	188.76 & NTT & EFOSC2 & 
Gr\#16 & 2700 & 1.0 &  13.4 &  6000--9900\\
2018 Jan. 16	& 58134.268 & 189.66	& VLT  & X-Shooter & UVB/VIS/NIR & 3600/3400/3680 & 0.9/0.9/1.0 & 1/1.1/3.3 & 3000--5560/5450--10,200/10,000--20,600\\
2018 Jan. 19	& 58137.305 & 192.40	& VLT  & X-Shooter & UVB/VIS/NIR & 3600/3400/3680 & 0.9/0.9/1.0 & 1/1.1/3.3 & 3000--5560/5450--10,200/10,000--20,600\\
2018 Feb. 14 & 58163.277 &	215.83 & NTT & EFOSC2 & 
Gr\#11 & 8100 & 1.0 &  13.8 &  3400--7400\\
2018 Feb. 18 & 58167.245 &	219.41 & NTT & EFOSC2 & 
Gr\#16 & 5400 & 1.0 &  13.4 &  6000--9900\\
2018 Apr. 9 & 58217.065 &	264.35 & NTT & EFOSC2 & 
Gr\#13 & 2700 & 1.0 & 18.2  &  3650--9200\\
\hline 
\end{tabular}
$^a$WiFeS is an integral field unit (IFU) with 25 slitlets that are $1''$ wide and $38''$ long. Resolution is measured from the night-sky lines.
\end{minipage}
\end{sidewaystable}

\clearpage

\appendix
One machine readable table: 
Pseudobolometric luminosity and photometry of SN~2017ens (UV through NIR). The magnitudes without uncertainties are 3$\sigma$ detection limits. ATLAS {\it cyan} and {\it orange} bands are also converted to the SDSS system using our spectra (for prediscovery detection limits we adopt the first-epoch passband correction value).

\begin{table*}
 \begin{minipage}{190mm}
 \centering
  \caption{Optical photometry of {\it griz} bands of SN~2017ens. The ``$>$'' denotes 3$\sigma$ detection limits. }
\label{tab:sn_phot_opi}
\begin{tabular}[t]{lllllllllllll}
\hline
UT Date & MJD & Phase & {\it g} (error) & {\it r} (error)& {\it i} (error) & {\it z} (error) & Telescope/Instrument \\
\hline
2017 June 8 		&57912.992&	-9.94	&17.57	(0.05)	&17.96	(0.05)	&18.24	(0.04)	&18.35	(0.06)	& GROND \\
2017 June 10 		&57914.026&	-9.01	&17.53	(0.03)	&17.89	(0.05)	&18.16	(0.04)	&18.28	(0.05)	& GROND \\
2017 June 13 		&57917.011&	-6.31	&17.41	(0.08)	&17.73	(0.04)	&18.02	(0.06)	&18.09	(0.06)	& GROND \\
2017 June 20 		&57924.011&	0.00	&17.33	(0.06)	&17.54	(0.04)	&-	&17.81	(0.06)	& GROND \\
2017 June 25 	&57929.465	&4.92	&17.50	(0.04)	&17.58	(0.03)	&17.67	(0.05) & - & LCO \\
2017 June 29		&57933.058&	8.16	&17.47	(0.01)	&17.59	(0.02)	&17.66	(0.06)	&17.70	(0.03)	& GROND \\
2017 July 2	&57937.012	&11.73	&17.63	(0.04)	&17.75	(0.07)	&17.76	(0.06) & - & LCO \\
2017 July 3		&57937.980&	12.60	&17.63	(0.07)	&17.65	(0.06)	&17.70	(0.07)	&17.74	(0.06)	& GROND \\
2017 July 3  &57938.031	&12.65	&17.78	(0.05)	&17.63	(0.05)	&17.64	(0.06) & - & LCO \\
2017 July 4  &57938.428	&13.00	&17.76	(0.08)	&17.74	(0.07)	&17.50	(0.08) & - & LCO \\
2017 July 7  &57941.345	&15.64	&17.85	(0.06)	&17.71	(0.06)	&17.73	(0.05) & - & LCO \\
2017 July 7  &57941.994	&16.22	&17.87	(0.04)	&17.82	(0.04)	&17.77	(0.04) & - & LCO \\
2017 July 10  &57945.007	&18.94	&17.99	(0.03)	&17.85	(0.03)	&17.86	(0.04) & - & LCO \\
2017 July 12		&57946.033&	19.86	&17.98	(0.03)	&17.89	(0.05)	&17.88	(0.05)	&17.95	(0.04)	& GROND \\
2017 July 14		&57948.027&	21.66	&18.07	(0.09)	&17.94	(0.03)	&17.95	(0.04)	&17.94	(0.07)	& GROND \\
2017 July 14	&57948.984	&22.53	&18.16	(0.07)	&17.90	(0.06)	&17.90	(0.06) & - & LCO \\
2017 July 16		&57950.995&	24.34	&18.20	(0.01)	&18.03	(0.03)	&17.97	(0.06)	&18.00	(0.05)	& GROND \\
2017 July 19		&57953.983&	27.04	&18.44	(0.07)	&18.21	(0.06)	&18.15	(0.10)	&18.19	(0.04)	& GROND \\
2017 July 20  &57954.967	&27.92	& -	&18.21	(0.08)	&18.12	(0.10) & - & LCO \\
2017 July 21		&57955.993&	28.85	&18.50	(0.05)	&18.28	(0.05)	&18.19	(0.06)	&18.23	(0.03)	& GROND \\
2017 July 24		&57958.984&	31.55	&18.61	(0.05)	&18.39	(0.05)	&18.25	(0.05)	&18.27	(0.04)	& GROND \\
2017 July 27	&57961.979&	34.25	&18.78	(0.04)	&18.54	(0.05)	&18.42	(0.06)	&18.42	(0.05)	& GROND \\
2017 July 31	&57965.983&	37.86	&18.96	(0.06)	&18.62	(0.06)	&18.51	(0.07)	&18.51	(0.06)	& GROND \\
2017 August 1		&57966.979&	38.76	&18.94	(0.04)	&18.66	(0.05)	&18.52	(0.06)	&18.49	(0.05)	& GROND \\
2017 August 9		&57974.968&	45.97	&19.20	(0.05)	&18.96	(0.23)	&18.73	(0.07)	&18.79	(0.07)	& GROND \\
2017 December 16	&58103.300&	161.73	&20.60	(0.05)	&20.62	(0.05)	&20.13	(0.07)	&20.15	(0.10)	& GROND \\
2017 December 22	&58109.303&	167.14	&20.58	(0.05)	&20.65	(0.06)	&20.11	(0.07)	&20.31	(0.09)	& GROND \\
2017 December 27	&58114.304&	171.65	&20.78	(0.06)	&20.99	(0.05)	&20.25	(0.06)	&20.51	(0.06)	& GROND \\
2018 January 4	&58122.318&	178.88	&20.91	(0.08)	&21.02	(0.06)	&20.26	(0.08)	&20.64	(0.07)	& GROND \\
2018 January 6		&58124.304&	180.67	&20.86	(0.06)	&20.89	(0.07)	&20.19	(0.05)	&20.53	(0.07)	& GROND \\
2018 January 12	&58130.266&	186.05	&21.01	(0.06)	&20.97	(0.06)	&20.37	(0.10)	&20.50	(0.06)	& GROND \\
2018 January 15	&58133.263&	188.75	&20.80	(0.05)	&20.83	(0.05)	&20.26	(0.06)	&20.41	(0.06)	& GROND \\
2018 January 18	&58136.303&	191.50	&20.91	(0.08)	&20.96	(0.05)	&20.31	(0.08)	&20.58	(0.05)	& GROND \\
2018 January 22	&58140.283&	195.09	&20.84	(0.07)	&20.98	(0.04)	&20.33	(0.09)	&20.61	(0.06)	& GROND \\
2018 January 24	&58142.282&	196.89	&20.82	(0.06)	&20.92	(0.05)	&20.27	(0.08)	&20.53	(0.06)	& GROND \\
2018 January 27	&58145.319&	199.63	&20.98	(0.06)	&21.10	(0.05)	&20.36	(0.09)	&20.73	(0.06)	& GROND \\
2018 February 1	&58150.333&	204.15	&20.92	(0.17)	&21.26	(0.14)	&20.63	(0.13)	&20.63	(0.16)	& GROND \\
2018 February 12	&58161.258&	214.01	&21.04	(0.05)	&20.94	(0.04)	&20.28	(0.09)	&20.61	(0.08)	& GROND \\
2018 February 20	&58169.268&	221.23	&20.93	(0.05)	&21.01	(0.05)	&20.34	(0.08)	&20.63	(0.07)	& GROND \\
2018 February 27	&58176.338		&227.61	&20.96	(0.08)	&21		(0.04)	&20.3	(0.05)	&20.83	(0.05)	& GROND\\
2018 March 19		&58196.179	&245.51	&21.23	(0.08)	&21.37	(0.05)	&20.47	(0.09)	&21.01	(0.06)	 & GROND\\
2018 March 25		&58202.285	&251.01	&21.05	(0.07)	&21.23	(0.06)	&20.36	(0.09)	&20.79	(0.07)		& GROND \\
2018 April 6		& 58214.119	&261.69	&21.05	(0.05)	&21.11	(0.06)	&20.41	(0.09)	&20.66	(0.08)		& GROND\\
2018 April 18		&58226.095	&272.49	&21.07	(0.07)	&21.02	(0.04)	&20.26	(0.05)	&20.72	(0.09)		& GROND\\
\hline 
\end{tabular}
\end{minipage}
\end{table*}

\begin{table*}
 \begin{minipage}{190mm}
 \centering
  \caption{Optical photometry of SN~2017ens in ATLAS $cyan$ and $orange$ bands.$^a$}
\label{tab:sn_phot_atlas}
\begin{tabular}[t]{lllllllllllll}
\hline
UT Date & MJD & Phase & {\it c} (error) & {\it o} (error)& {\it c $\rightarrow$ g} & {\it o $\rightarrow$ r} \\
\hline
2017 May 10		&57883.428&	-36.61	&- &$>19.03$			&- &$>19.08$	\\
2017 May 10		&57883.443&	-36.59	&- &$>19.13$			&- &$>19.18$	\\
2017 May 10		&57883.459&	-36.58	&- &$>18.89$			&- &$>18.94$	\\
2017 May 10		&57883.478&	-36.56	&- &$>18.76$			&- &$>18.81$	\\
2017 May 11		&57884.365&	-35.76	&- &$>19.58$			&- &$>19.63$	\\
2017 May 11		&57884.401&	-35.73	&- &$>19.35$			&- &$>19.4	$\\
2017 May 11		&57884.416&	-35.72	&- &$>19.32$			&- &$>19.37$	\\
2017 May 13		&57886.285&	-34.03	&- &$>19.69$			&- &$>19.74$	\\
2017 May 13		&57886.299&	-34.02	&- &$>19.66$			&- &$>19.71$	\\
2017 May 13		&57886.314&	-34.00	&- &$>19.53$			&- &$>19.58$	\\
2017 May 13		&57886.328&	-33.99	&- &$>19.53$			&- &$>19.58$	\\
2017 May 19		&57892.370&	-28.54	&- &$>20.31$			&- &$>20.36$	\\
2017 May 19		&57892.399&	-28.52	&- &$>20.23$			&- &$>20.28$	\\
2017 May 19		&57892.427&	-28.49	&- &$>20.25$			&- &$>20.3	$\\
2017 May 20		&57893.299&	-27.70	&- &$>20.22$			&- &$>20.27$	\\
2017 May 20		&57893.317&	-27.69	&- &$>20.18$			&- &$>20.23$	\\
2017 May 20		&57893.350&	-27.66	&- &$>19.70$			&- &$>19.75$	\\
2017 May 23		&57896.373&	-24.93	& $>20.06$ & -			& $>20.00$ & -	\\
2017 May 24		&57897.289&	-24.10	&- &$>20.06$			&- &$>20.11$	\\
2018 May 24		&57897.306&	-24.09	&- &$>20.03$			&- &$>20.08$	\\
2019 May 24		&57897.323&	-24.07	&- &$>20.06$			&- &$>20.11$	\\
2017 May 24		&57897.340&	-24.06	&- &$>20.14$			&- &$>20.19$	\\
2017 May 25		&57898.332&	-23.16	&- &$>20.18$			&- &$>20.23$	\\
2017 May 27		&57900.346&	-24.93	& $>20.36$ & -			& $>20.30$ & -	\\
2017 May 27		&57900.381&	-24.32	& $>20.34$ & -			& $>20.28$ & -	\\
2017 May 27		&57900.416&	-24.28	& $>20.35$ & -			& $>20.29$ & -	\\
2017 May 29		&57902.268&	-19.61	&- &$>20.29$			&- &$>20.34$	\\
2017 May 29		&57902.301&	-19.58	&- &$>20.12$			&- &$>20.17$	\\
2017 May 29		&57902.318&	-19.57	&- &$>20.13$			&- &$>20.18$	\\
2017 May 31		&57904.286&	-17.79	& $>19.65$ & -			& $>19.59$ & -	\\
2017 June  1	&57905.257&	-16.92	&- &$>19.81$			&- &$>19.86$	\\
2017 June  1	&57905.271&	-16.90	&- &$>19.66$			&- &$>19.71$	\\
2017 June  2	&57906.257&	-16.01	&- &$>19.40$			&- &$>19.45$	\\
2017 June  4	&57908.326&	-14.15	&- &$>17.81$			&- &$>17.86$	\\
2017 June  4	&57908.394&	-14.09	&- &$>17.88$			&- &$>17.93$	\\
2017 June  5 	&57909.274&	-13.29	&- &$18.62	\pm0.12$	&- &$18.67$\\
2017 June  6 	&57910.275&	-12.39	&- &$18.6	\pm0.10$	&- &$18.65$\\
2017 June  9 	&57913.276&	-9.68	&- &$18.08	\pm0.10$	&- &$18.13$\\
2017 June 10 	&57914.267&	-8.79	&- &$17.86	\pm0.08$	&- &$17.91$\\
2017 June 12 	&57916.327&	-6.93	&- &$17.94	\pm0.05$	&- &$17.99$\\
2017 June 15 	&57919.320&	-4.23	&- &$17.78	\pm0.06$	&- &$17.83$\\
2017 June 16 	&57920.316&	-3.33	&- &$17.68	\pm0.02$	&- &$17.73$\\
2017 June 19 	&57923.327&	-0.62	&- &$17.69	\pm0.16$	&- &$17.74$\\
2017 June 20 	&57924.303&	0.26	&- &$17.57	\pm0.04$	&- &$17.62$\\
2017 June 23 	&57927.304&	2.97	&- &$17.62	\pm0.03$	&- &$17.67$\\
2017 June 23 	&57927.322&	2.99	&- &$17.64	\pm0.05$	&- &$17.69$\\
2017 June 27 	&57931.297&	6.57	&- &$17.56	\pm0.03$	&- &$17.61$\\
2017 July  5 	&57939.275&	13.77	&- &$17.67	\pm0.06$	&- &$17.77$\\
2017 July  8 	&57942.273&	16.47	&- &$17.78	\pm0.07$	&- &$17.88$\\
2017 July  9 	&57943.255&	17.36	&- &$17.75	\pm0.08$	&- &$17.85$\\
2017 July 10 	&57944.261&	18.27	&- &$17.81	\pm0.06$	&- &$17.91$\\
2017 July 11 	&57945.263&	19.17	&- &$17.9	\pm0.04$	&- &$18.00$\\
2017 July 15 	&57949.261&	22.78	&- &$17.96	\pm0.05$	&- &$18.06$\\
2017 July 17 	&57951.260&	24.58	&- &$18.09	\pm0.05$	&- &$18.29$\\
2017 July 21 	&57955.258&	28.19	&- &$18.26	\pm0.08$	&- &$18.46$\\
\hline 
\end{tabular}
\end{minipage}
$^a$We also convert these to the SDSS system using our spectra. 

For prediscovery detection limits we adopt the first-epoch S-correction value.
\end{table*}

\begin{table*}
 \begin{minipage}{190mm}
 \centering
  \caption{Optical photometry of SN~2017ens using Lulin SLT telescope. }
\label{tab:sn_phot_slt}
\begin{tabular}[t]{lllllllll}
\hline
UT Date & MJD & Phase & {\it B} & {\it V} & {\it R} & {\it I} \\
\hline
2017 June 23	&57927.572	&3.21	&17.55	(0.15)	&17.36	(0.19)	&17.30	(0.25)	&17.27	(0.26)	\\
2017 June 26	&57930.551	&5.90	&17.77	(0.13)	&17.36	(0.25)	&17.29	(0.35)	&17.07	(0.30) \\
\hline 
\end{tabular}
\end{minipage}
\end{table*}

\begin{table*}
 \begin{minipage}{190mm}
 \centering
  \caption{Ultraviolet photometry of SN~2017ens in the {\it Swift} UVOT bands.}
\label{tab:sn_phot_uv}
\begin{tabular}[t]{lllllllll}
\hline
UT Date & MJD & Phase & {\it v} & {\it b} & {\it u} & {\it uvw1} & {\it uvm2} & {\it uvw2} \\
\hline
2017 June 28	&57932	&7.21	&17.58	(0.15)	&17.59	(0.09)	&17.91	(0.08)	&18.48	(0.07)	&18.72	(0.06)	&18.96	(0.08)\\
2017 July  3	&57937	&11.72	&17.50	(0.16)	&17.88	(0.11)	&18.00	(0.09)	&18.78	(0.09)	&19.01	(0.07)	&19.20	(0.08)\\
2017 July 25	&57959	&31.56	&18.85	(0.37)	&18.85	(0.18)	&19.13	(0.13)	&20.10	(0.13)	&20.33	(0.10)	&20.53	(0.11)\\
2018 January 22	&58140	&194.83	& -	& -	& -	& $>21.85$	& $>22.61$	&22.36	(0.26)\\
2018 January 30	&58150	&203.85	& -	& -	&21.18	(0.13)	& -	& -	&22.42	(0.21)\\
\hline 
\end{tabular}
\end{minipage}
\end{table*}

\begin{table*}
 \begin{minipage}{190mm}
 \centering
  \caption{NIR photometry of SN~2017ens in the GROND NIR bands.}
\label{tab:sn_phot_nir}
\begin{tabular}[t]{llllllll}
\hline
UT Date & MJD & Phase & {\it J} & {\it H} & {\it K$_{s}$} \\
\hline
2017 June 8 	&57912.992	&-9.94	&18.80	(0.12)	&19.14	(0.19)	&$>19.52$	 \\
2017 June 10	&57914.026	&-9.01	&18.82	(0.11)	&19.31	(0.18)	&$>19.20$	 \\
2017 June 13	&57917.011	&-6.31	&18.52	(0.10)	&18.89	(0.13)	&19.18	(0.27) \\
2017 June 20	&57924.011	&0.00	&18.38	(0.10)	&18.59	(0.13)	& - \\
2017 June 29	&57933.058	&8.16	&18.18	(0.11)	&18.52	(0.14)	&18.64	(0.20) \\
2017 July 3 	&57937.980	&12.60	&18.16	(0.16)	&18.31	(0.20)	& - \\
2017 July 12	&57946.033	&19.86	&18.37	(0.10)	&18.61	(0.15)	&18.85	(0.21) \\
2017 July 14	&57948.027	&21.66	&18.19	(0.12)	&18.66	(0.18)	&$>18.85$	 \\
2017 July 16	&57950.995	&24.34	&18.48	(0.15)	&18.83	(0.26)	&$>18.93$	 \\
2017 July 19	&57953.983	&27.04	&18.42	(0.12)	&18.89	(0.14)	&19.10	(0.24) \\
2017 July 21	&57955.993	&28.85	&18.52	(0.10)	&18.74	(0.14)	&19.10	(0.18) \\
2017 July 24	&57958.984	&31.55	&18.63	(0.10)	&19.10	(0.19)	&$>19.42$	 \\
2017 July 27	&57961.979	&34.25	&18.70	(0.11)	&19.06	(0.16)	&$>19.48$	 \\
2017 July 31	&57965.983	&37.86	&18.79	(0.16)	&$>19.51$	&$>19.29$	 \\
2017 August 1 	&57966.979	&38.76	&18.87	(0.18)	&$>19.47$	&$>19.26$	 \\
2017 August 9 	&57974.968	&45.97	&19.02	(0.26)	&$>19.18$	& -\\
2017 December 16  &58103.300	&161.73	&$>20.91$	&19.64	(0.21)	&$>19.59$\\
2017 December 22  &58109.303	&167.14	&20.03	(0.18)	&$>20.37$	& - \\
2017 December 27  &58114.304	&171.65	&20.42	(0.15)	&20.47	(0.20)	&19.86	(0.23) \\
2018 January 4 	&58122.318	&178.88	&20.42	(0.21)	&20.30	(0.26)	&$>19.82$ \\
2018 January 6 	&58124.304	&180.67	&20.35	(0.15)	&20.72	(0.28)	&19.76	(0.26) \\
2018 January 12 &58130.266	&186.05	&20.77	(0.22)	&19.93	(0.18)	& - \\
2018 January 15 &58133.263	&188.75	&20.35	(0.17)	&20.18	(0.21)	&19.73	(0.22) \\
2018 January 18 &58136.303	&191.50	&20.34	(0.16)	&20.20	(0.21)	& - \\
2018 January 22 &58140.283	&195.09	&20.33	(0.17)	&20.35	(0.25)	&19.95	(0.30) \\
2018 January 24 &58142.282	&196.89	&20.63	(0.23)	&20.24	(0.25)	&19.60	(0.24) \\
2018 January 27 &58145.319	&199.63	&20.49	(0.18)	&20.18	(0.18)	&$>19.60$ \\
2018 February 1 &58150.333	&204.15	&$>20.53$	&$>20.26$	&$>18.97$ \\
2018 February 12 	&58161.258&	214.01		&20.43	(0.14)	&20.24	(0.17)	&$>19.85$ \\
2018 February 18 	&58167.310&	219.46		&20.39	(0.17)	&20.48	(0.24)	&$>19.82$ \\
2018 February 20 	&58169.268&	221.23		&20.57	(0.24)	&$>20.70$	&$>19.93$ \\
2018 February 27	&58176.338		&227.61	&20.32	(0.15)	&20.58	(0.22)	&19.52		(0.26) \\
2018 March 7		&58184.248	&234.74	&20.43	(0.19)	&20.39	(0.24)	&$>19.981$ \\
2018 March 19		&58196.179	&245.51	&20.46	(0.16)	&20.25	(0.17)	&19.63		(0.24) \\
2018 March 25		&58202.285	&251.01	&20.53	(0.18)	&20.43	(0.23)	&$>19.92$ \\
2018 April 6		& 58214.119	&261.69	&20.66	(0.24)	&20.53	(0.32)	&$>19.99$ \\
2018 April 18		&58226.095	&272.49	&$>21.07$		&$>20.61$		&$>19.44$ \\
\hline 
\end{tabular}
\end{minipage}
\end{table*}

\begin{table}
 \begin{minipage}{83mm}
 \centering
  \caption{Pseudobolometric luminosity of SN~2017ens (UV through NIR).} 
\label{tab:bol_flux}
\begin{tabular}{c c c c}
\hline
Phase & $\mathrm{log}\,L_{\rm bol}$ & Phase & $\mathrm{log}\,L_{\rm bol}$ \\  
(day) & (erg\,s$^{-1}$) & (day) & (erg\,s$^{-1}$)   \\  
\hline
-13.30 	&$43.45  \pm0.08$& 19.86 	&$43.61  \pm0.01$\\
-12.39 	&$43.45  \pm0.08$& 21.66 	&$43.59  \pm0.01$\\
-9.94	&$43.65  \pm0.01$& 22.53	&$43.58  \pm0.01$\\
-9.69	&$43.65  \pm0.01$& 22.78 	&$43.57  \pm0.01$\\
-9.01	&$43.67  \pm0.01$& 24.34 	&$43.55  \pm0.01$\\
-8.80	&$43.68  \pm0.01$& 24.58 	&$43.54  \pm0.01$\\
-6.94	&$43.70  \pm0.01$& 27.04 	&$43.48  \pm0.01$\\
-6.31	&$43.73  \pm0.01$& 27.92	&$43.47  \pm0.01$\\
-4.23	&$43.74  \pm0.01$& 28.18 	&$43.47  \pm0.02$\\
-3.34	&$43.75  \pm0.01$& 28.85	&$43.46  \pm0.02$\\
-0.62	&$43.77  \pm0.01$& 31.55	&$43.42  \pm0.02$\\
0.00	&$43.78  \pm0.01$& 34.25	&$43.36  \pm0.02$\\
0.26	&$43.78  \pm0.01$& 37.86	&$43.31  \pm0.02$\\
2.97	&$43.76  \pm0.01$& 38.76	&$43.31  \pm0.02$\\
4.92	&$43.76  \pm0.01$& 45.97	&$43.23  \pm0.03$\\
6.57	&$43.77  \pm0.01$& 161.73	&$42.65  \pm0.03$\\
8.16 	&$43.76  \pm0.00$& 167.14	&$42.69  \pm0.03$\\
11.73	&$43.71  \pm0.01$& 171.65	&$42.56  \pm0.03$\\
12.60 	&$43.71  \pm0.01$& 178.88	&$42.53  \pm0.03$\\
12.65	&$43.69  \pm0.01$& 180.67	&$42.56  \pm0.03$\\
13.00	&$43.70  \pm0.01$& 186.05	&$42.54  \pm0.03$\\
13.76 	&$43.69  \pm0.01$& 188.75	&$42.59  \pm0.03$\\
15.64	&$43.66  \pm0.01$& 191.50	&$42.56  \pm0.04$\\
16.22	&$43.65  \pm0.01$& 195.09	&$42.54  \pm0.04$\\
16.47 	&$43.65  \pm0.01$& 196.89	&$42.54  \pm0.04$\\
17.35 	&$43.65  \pm0.01$& 199.63	&$42.50  \pm0.04$\\
18.27 	&$43.64  \pm0.01$& 204.15	&$42.49  \pm0.05$\\
18.94	&$43.62  \pm0.01$& 214.01	&$42.52  \pm0.04$\\
19.17 	&$43.62  \pm0.01$& 221.23	&$42.52  \pm0.04$\\
\hline 
\end{tabular}
\end{minipage}
\end{table}

\end{document}